# Engineering THz-frequency light generation, detection and manipulation through graphene

**Miriam S. Vitiello and Leonardo Viti**
[1]*NEST, CNR-NANO and Scuola Normale Superiore, 56127, Pisa, Italy*

## Abstract

Graphene has been one of the most investigated materials in the last decade. Its unique optoelectronic properties have indeed raised it to an ideal and revolutionary candidate for the development of entirely novel technologies across the whole electromagnetic spectrum, from the microwaves to the x-rays, even crossing domain of intense application relevance, as terahertz (THz) frequencies.
Owing to its exceptionally high tensile strength, electrical conductivity, transparency, ultra-fast carrier dynamics, non-linear optical response to intense fields, electrical tunability and ease of integration with semiconductor materials, graphene is a key disruptor for the engineering of generation, manipulation, and detection technologies with *ad-hoc* properties, conceived from scratch.
In this review, we elucidate the fundamental properties of graphene, with an emphasis on its transport, electronic, ultrafast and non-linear interactions, and explore its enormous technological potential of integration with a diverse array of material platforms. We start with a concise introduction to graphene physics, followed by the most remarkable technological developments of graphene-based photodetectors, modulators, and sources in the 1-10 THz frequency range. As such, this review aims to serve as a valuable resource for a broad audience, ranging from novices to experts, who are keen to explore graphene physics for conceiving and realizing micro- and nano-scale devices and systems in the far infrared. This would allow addressing the present challenging application needs in quantum science, wireless communications, ultrafast science, plasmonics and nanophotonics.

## 1. Introduction

In the electromagnetic spectrum, the frequency range 0.1–10 THz (wavelength of 30 $\mu$m – 3 mm) identifies the spectral domain that crosses the infrared and the microwave regions, i.e. the areas commonly of interest in photonics and electronics.

The interaction between terahertz (THz) frequency light and matter was a mostly neglected research area for a long time, until ground-breaking advances in the technology to generate, detect and manipulate THz frequency beams were achieved. The most relevant milestones include the development of ultrafast switches [1, 2], and photoconductive antennas [3], the demonstration of optical rectification [4] and free-space electro-optical sensing [5] and the invention of time-domain spectroscopy (TDS) [6], in all cases resulting in coherent detection schemes requiring complex, massive and expensive setup, that hinders real-time in situ applications in exciting fields as wireless communications [7, 8], non-destructive security inspection [9], biomedicine [10-13], food and agricultural product safety detection [14, 15], global environmental monitoring [16-18], industrial applications [19], defence [20, 21], and so on.

In the recent past, two-dimensional (2D) materials attracted broad interest as new material platforms across the THz. Graphene has been one of the most intensively investigated 2D materials. Owing to its gapless nature [22] (Figure 1a), it indeed exhibits an unique set of properties, such as low electronic heat capacity of the 2D massless electron gas, ultrafast (picosecond) electron cooling dynamics, broadband absorption, and electrical tunability [50], that endow graphene with a strong, nonlinear, ultrafast, and adjustable response to THz frequency radiation. As a consequence, a variety of graphene devices and systems operating at THz and sub-THz frequencies has been reported, including fast photodetectors [23-27], amplitude, phase, frequency and polarization modulators [28-30], sources, plasmonic devices [31], and ultrafast lasers [32, 33].

Besides these exceptional transport and optical properties, high-quality graphene can be nowadays also synthesized on wafer-scale [34-42], becoming compatible with large-scale production and with the silicon pilot-line, motivating incredible research efforts in devices engineering. As an example, broadband image sensor arrays have been developed by integrating large-area graphene grown by chemical vapor deposition (CVD) with complementary metal-oxide semiconductor (CMOS) technology. These arrays can operate in a wide range of wavelengths, spanning from ultraviolet (4−400 nm) to short-wave infrared (SWIR, 900−2000 nm) [43]. The integration with various material platforms has further enabled new regime of operation. For example, graphene field effect transistors (GFETs) fabricated on a pyroelectric substrate have been demonstrated to efficiently operate as uncooled mid-infrared pyroelectric bolometers with impressive temperature coefficients of resistance (~900% $K^{-1}$), capable of resolving temperature variations as small as 15 μK [44]. Furthermore, the integration of single crystal CVD graphene with photonic waveguides has recently enabled the realization of optoelectronic devices for communications: ultrafast, zero-dark-current photodetectors (PDs) operating in the telecom band [45], and graphene electro-absorption modulators with data rates up to 20 Gbps realized on wafer-scale matrices comprising up to ~12000 individual elements [46].

In addition, graphene also displays an extremely high room-temperature electron mobility (up to $2.5\times10^5$ $cm^2V^{-1}s^{-1}$) [47, 48] and a frequency independent absorption/layer of $\alpha\pi \approx 2.3$ %, being α the fine constant [49], for large enough photon energies, i.e. when $\hbar\omega \gg 2E_F$ (Figure 1b) and $\hbar\omega \gg 4k_BT$ [50], where ħ is the reduced Planck constant and $k_B$ is the Boltzmann constant. Due to the fundamental absence of Coulomb binding, electron transport is not affected by impurity trapping and other unscreened Coulomb potentials and is only limited by the momentum scattering. This property, combined with the large optical phonon energy (≈200 meV), and the nonpolar nature of its lattice, results in carrier mobilities that can even exceed 100.000 $cm^2V^{-1}s^{-1}$[51-53].

Furthermore, its linear momentum and energy dispersion can allow triggering numerous phenomena like anomalous quantum Hall effect [54], Klein tunneling [55-57], and tunable interband transitions [58]. Its electrical properties can be modified through doping or electrostatic gating, making it ideal for tunable device architectures. Similarly, light absorption can be significantly enhanced via the excitation of surface plasmon resonance from the ultraviolet to the microwave band [59-62]. Finally, single layer graphene (SLG) is a highly nonlinear material ($\chi^{(3)}$~$10^{-9}$ $m^2/V^2$ in the far-infrared [63]) and it is thus ideal for integration into nonlinear optical devices [49, 64, 65]. The lack of bandgap and the superb electrical tunability, can enable shifting the Fermi energy $E_F$, and inducing Pauli blocking of the optical transitions [32, 66, 67]. Graphene also exhibits saturable absorption across a broad range of frequencies [68-70], large nonlinear optical response [50, 63] with a transmission modulation of ~50% per layer [70, 71], ultrafast photoexcitation dynamics and recovery time [50, 72-74], high chemical and mechanical stability [75], large thermal [76] and optical threshold damage [77] and it is also an extremely efficient THz frequency multiplier, with a huge 3rd order non-linearity, that allowed for the direct generation of multiple harmonics in the 0.85-2.12 THz range [78,79].

In addition, graphene also displays a strong ambipolar field-effect when subjected to a gate voltage, that can lead to a precise control of the type and density of the free charge carriers (electrons or holes) in the respective bands [47]. Combined with the high electron mobility, this property made graphene a valuable platform for the development of ultrafast transistors operating at rates of 100s of GHz [80, 81], and for engineering ultrafast devices, modulators, integrated sources, and detectors operating across the far-infrared.

Table I summarizes the fundamental physical properties of graphene, which are relevant for the engineering of devices operating in the THz frequency range.

| Property | Symbol | Typical value for single layer graphene | Relevance for THz devices |
|---|---|---|---|
| Electron mobility at room temperature | $\mu$ | $10^4$ for CVD grown<br>$> 10^5$ for hBN encapsulated | High-current density,<br>Low device impedance |
| Seebeck coefficient | $S_b$ | $\sim$100 μV/K | Good photothermoelectric conversion in THz detectors |
| Electronic heat capacity | $c_e$ | $\sim$2000 $k_B \mu m^{-2}$ at room temperature | Large temperature gradients in the electronic distribution |
| Third-order Susceptibility | $\chi^{(3)}$ | $\chi^{(3)} \sim 10^{-9}$ m$^2$/V$^2$ | Strong nonlinear response |
| Carrier heating time | $\tau_h$ | < 100 fs | Ultrafast thermalization of the electronic population |
| Carrier cooling time | $\tau_c$ | $\sim$1-4 ps (limited by optical phonon scattering) | Fast recovery after electron heating |
| THz Absorbance | $A$ | $\sim$10% | Efficient and tunable light modulation |

**Table I.** Summary of the main physical parameters that identify graphene as a core material for the development of THz devices.

One core property of graphene, which is relevant to introduce for understanding most of the results presented in the present review, is its Fermi level $E_F$-dependent optical conductivity $\sigma(n)$, that comprises both interband, $\sigma_{inter}$, and intraband, $\sigma_{intra}$, terms. At THz frequencies, $\sigma_{inter}$ is more than three orders of magnitude lower than $\sigma_{intra}$ [82], and can be then safely neglected (Figure 1b). The total conductivity is therefore usually calculated using the Kubo formula [83]:

$$\sigma_{\text{intra}}(\nu, E_F, \Gamma_0, T) = \frac{-iD_0}{\pi} \frac{1}{(2\pi\nu + i\Gamma_0)} \left[1 + \frac{k_B T}{E_F} 2ln\left(e^{-E_F/k_B T} + 1\right)\right] \quad \text{(e1)}$$

where $D_0 = E_F e^2/\hbar^2$ is the linear Drude weight, $e$ is the electron charge and $T$ the temperature, $\Gamma_0=(ev_F^2)/(E_F \mu)$ is the scattering rate, sensitive to $E_F$ and to the carrier mobility $\mu$, and $v_F$ is the Fermi velocity. It is worth noting that equation (e1) can be simplified by neglecting the contribution in the square brackets, resulting in a simple Drude model.

In the presence of an intense excitation beam, the nonlinear response of SLG can be expressed using a field-dependent conductivity [84]:

$$\sigma_{\text{tot}}(\nu) = \sigma_{\text{intra}}(\nu) + |E_0(\nu)|^2 \times \sigma_3(\nu) \quad \text{(e2)}$$

where $E_0$ represents the electric field amplitude. The nonlinear term of the conductivity, $\sigma_3$, is composed of the 3rd harmonic and Kerr effect terms, whose numerical expressions, in the case of graphene shaped as a metamaterial or embedded in a metamaterial-like resonator, are [84]:

$$\sigma_3(\nu) = \eta[\sigma_{\text{Kerr}}(\nu)] \quad \text{(e3)}$$

$$\sigma_{\text{Kerr}}(\nu = \nu_{\text{eff}}) = \frac{i9e^6 v_F}{4\pi\hbar^4} \frac{D_{he}}{(2\pi\nu + i\Gamma_{he})(-2\pi\nu + i\Gamma_{he})(4\pi\nu + i\Gamma_{he})} \quad \text{(e4)}$$

Where $\eta$ is the filling factor of the metamaterial with respect to the surrounding medium. For metamaterial-like resonators with resonance frequency $\nu_0$, the frequency $\nu$ is replaced by $\nu_{\text{eff}}=(\nu^2-\nu_0^2)/\nu$. The $E_F$-dependent parameters $D_{he}$ and $\Gamma_{he}$ are the hot-electron Drude weight and scattering rate, respectively. These can be written as [85]:

$$D_{he} = D_0\left[1 - \frac{1}{6}\left(\frac{\pi k_B}{E_F}\right)^2 T_e^{\,2}\right] \quad \text{(e5)}$$

$$\Gamma_{he} = \Gamma_0\left[1 + \frac{1}{6}\left(\frac{\pi k_B}{E_F}\right)^2 T_e^{\,2}\right] \quad \text{(e6)}$$

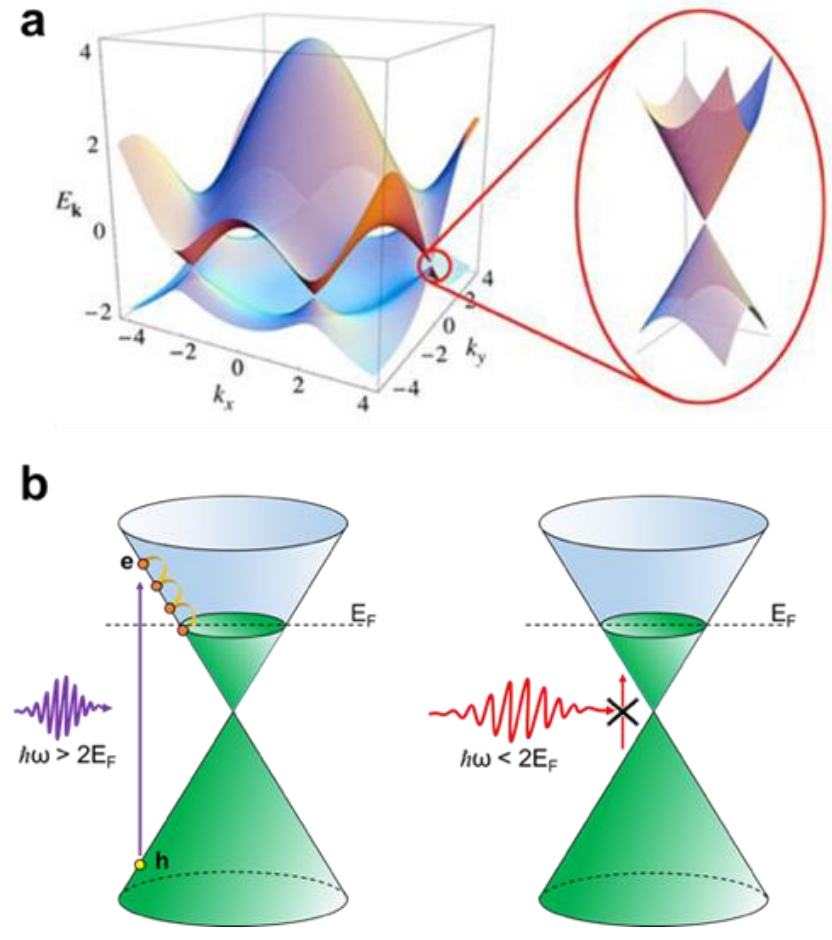

**Figure 1. Electronic band structure of graphene. (a)** The conduction and valence bandd intersect at six vertices of the hexagonal Brillouin zone, known as K points. The zoom shows that the dispersion relation near the K points is linear, resembling the energy spectrum of massless Dirac particles. Adapted from ref. [87]: reproduced with permission from Reviews of Modern Physics, 81, 109 (2009), Copyright 2009, American Physical Society. **(b)** Schematic of light-matter interaction in doped graphene ($E_F \neq 0$) for different photon energies ($\hbar\omega$). When $\hbar\omega > 2E_F$, interband transitions can occur, leading to the formation of electron-hole pairs. Importantly, under these conditions, the high-energy electron thermalizes, resulting in the generation of several *hot* electrons. Conversely, when $\hbar\omega < 2E_F$, interband absorption is hindered by Pauli blocking, and intraband conductivity dominates the photoresponse. When subjected to low-energy (THz) optical excitation, electrons near the Fermi surface are accelerated by the impinging field through a free-carrier absorption mechanism.

The model described by equations (e1-e6) predicts that for weak excitation and large doping, ($k_B T_e \ll E_F$), the hot-carrier conductivity is lower than the equilibrium conductivity, resulting in negative photoconductivity.

More details can be found in a recent topical review [50]. In the case of polycrystalline graphene, the Drude model (eq.1) should be replaced by the more general Drude-Smith model [86].

## 2.THz Photodetectors

In the past decades, THz-frequency photodetectors have witnessed a tremendous expansion. Different device architestures have been proposed and a plethora of target applications demonstrated, including imaging [88], security screening [89, 90], gas detection [91], astrophysical observation [92], biomedicine [93, 94], cultural heritage [95, 96] and high data-rate communications [97]. Commercially available technologies can be broadly categorized into two groups: cryogenically cooled ultra-sensitive devices and portable photodetectors that operate room-temperature. The first category includes bolometers based either on superconducting active elements [92, 98] or semiconductors (Si, Ge, InSb) [99], and gallium-doped germanium (Ge:Ga) photoconductive detectors [100, 101]. Cryogenically cooled photodetectors can achieve noise equivalent powers (NEP - optical power needed to reach unitary signal-to-noise ratio) $<10^{-14}$ $WHz^{-1/2}$ [99]. The second category includes single-pixel thermal detectors (Golay-cells, pyroelectrics, thermopiles) [102], and fast electronic rectifiers, such as Schottky diodes [103] and three-terminal field-effect transistors (FETs) based on complementary metal–oxide–semiconductor (CMOS) technology [104, 105]. In recent years, these latter, along with microbolometers and nanobolometers [106] and plasmonic photoconductive antennas [107], have been developed into large multi-pixel focal plane array systems, enabling a significant advancement in the evolution of real-time THz imaging systems.
Recent topical reviews have described the state-of-the-art of THz detectors [108] and the role of two dimensional materials in this technological context [109].

Over the past decade, owing to its ultra-broadband frequency response, significant efforts have been dedicated to the development of graphene-based photodetectors [25, 74]. Importantly, different physical mechanisms can underpin photodetection at THz frequencies in graphene [25]. The main detection mechanisms are the photo-conductive (PC) [110], photo-thermoelectric (PTE) [111, 112], photo-bolometric (PB) [113], plasma-wave (PW) [24, 114] and ballistic [115] rectification. These strongly depend on the energy of the impinging photons, on the operating temperature and on the specific detector design [25]. As a general remark on this latter, owing to the typically low absorption cross-section of few-layer thin materials, graphene-based THz detectors usually comprise antenna-integrated devices, where the antenna/resonator design enables to select the area (~0.1–1 $\mu m^2$) within graphene where photo-excitation occurs.

Research in the field is presently evolving across two main directions. On one hand, there is a major need of detectors which are highly sensitive (i.e., have a low noise-equivalent power, NEP) and that simultaneously operate at room temperature (RT), exhibit a fast photoresponse, display a broad dynamic range (the range between the lowest and highest measurable incident light power), work over a broad range of THz frequencies, and are possibly suitable for up-scaling to multi-pixel architectures [27, 116, 117]. On the other hand, the properties of graphene and its heterostructures are actively investigated for realizing ultra-sensitive quantum detectors, operating at cryogenic temperatures [118-120].

### 2.1 Room Temperature Terahertz Detectors

As discussed in the introduction, the absorption of THz radiation occurs primarily through intraband processes, in which the equilibrium thermal distribution of mobile carriers is excited and accelerated by an incident terahertz wave [50]. This excitation can lead to a thermal or a plasmonic response, or even to a combination of both [114]. Thermal effects, such as PTE and PB, are driven, in graphene, by the excess carrier temperature $\Delta T_e$, which can be large (up to ~1000 K [63]) thanks to the record small electronic heat capacity ($c_e$~2000 $k_B\mu m^{-2}$ at 300 K, where $k_B$ is the Boltzmann constant [121]) and to the weak (slow, $\tau_c$~ ps [122]) electron-phonon coupling compared to the strong (fast, $\tau_h$~ 10s fs [69, 123]) electron-electron coupling, which leaves photoexcited carriers in a *quasi*-equilibrium state [124] where their temperature ($T_e$) is significantly larger than the lattice and heat-sink temperatures. Importantly, the fast carrier cooling dynamics allows detection speed with timescales of few ps [125]. Thanks to these features, the PTE mechanism is typically dominating THz photoresponse of graphene-based devices at room-temperature.

The PTE mechanism relies on the Seebeck effect [50], in which a photovoltage (or photocurrent) is generated when light induces carrier heating at the junction between regions with different Seebeck coefficients ($S_b$) [126]. In order to maximize the PTE, $S_b$ must therefore be controlled. In graphene, $S_b$ is usually described theoretically by the Boltzmann theory [116]. In the degenerate limit ($k_BT$<< $E_F$), the linear Boltzmann equation results in an intuitive expression for $S_b$, given by the Mott equation, which relates $S_b$ to the *dc* conductivity of graphene and has been verified by many experiments [127, 128]: $S_{Mott} = -eL_0T \times \sigma^{-1}(\partial\sigma/\partial E_F)$, where $L_0 = (\pi k_B)^2/(3e^2)$ is the Lorenz number, and $E_F$ is the Fermi energy. From the Mott equation it is thus evident that a convenient way to modify $S_b$ is to modify graphene conductivity. In practical devices, this is typically achieved by tuning $E_F$ via electrostatic gating in graphene-based field effect transistors (GFETs). By controlling $E_F$ within specific areas of a graphene channel, p-n, p-p', or n-n' junctions can be established. Whenever these junctions with different $S_b$ are spatially overlapped with a light-induced temperature gradient (Figure 2a), e.g. when the junction is located at the center of a THz antenna, PTE conversion occurs.

This kind of mechanism has been demonstrated in high-quality hBN-encapsulated GFETs [26, 112, 129], obtaining state-of-the-art performances with NEP < 100 $pWHz^{-1/2}$ and response time ($\tau$) <1 ns (setup-limited). More recently, extensive research has been devoted to the improvement of the technological maturity of this kind of detectors, turning away from high-quality ($\mu$ ~$10^5$ $cm^2V^{-1}s^{-1}$ at room temperature) layered material heterostructures (LMHs), obtained by mechanical exfoliation and encapsulation of SLG in hexagonal boron nitride (hBN) [130], and moving towards wafer-scale SLG grown by chemical vapour deposition [131]. A major challenge in this transition is represented by the degradation of Sb when the material quality is lowered [132]: whereas $S_b$ up to 180 $\mu VK^{-1}$ has been obtained in hBN-encapsulated graphene [133], $S_b$ is typically lower (~10-50 $\mu VK^{-1}$) [134] in CVD-grown SLG. A possible route to overcome this issue relies on the choice of a suitable dielectric material surrounding graphene. For example, CVD graphene transferred on a tempered polymeric dielectric, poly(vinyl alcohol) (PVA), maintains its crystal quality, enabling large $S_b$ up to 140 $\mu VK^{-1}$ [45]. More recently, $S_b$ ~100 $\mu VK^{-1}$ has been achieved by using large-area graphene encapsulated in large-area hBN [116]. Sub-THz [117, 135, 136] and THz [27, 116] PTE detection has been demonstrated with scalable graphene-based architectures, reaching NEP~100 pWHz$^{-1/2}$ and response times of few ns. These performance, the possibility to integrate CVD-based GFETs with silicon platforms and the zero-bias operation make PTE GFETs very promising for future implementation within THz imaging systems.

Graphene based THz bolometers can also be very fast at room temperature, since hot-carrier dynamics enable detection speed with timescales of a few tens of picoseconds [113]. Differently from a PTE converter, a hot-carrier bolometer detects incident radiation by measuring the light-induced temperature increase of the electronic bath, which results in a global change in conductance of a graphene channel (Figure 2b) [50]. Thus, unlike PTE photodetectors, bolometers generally need a bias through the graphene channel in order to produce a photo-current [137]. This feature typically increases the noise figure of a detector [138] and increases its power consumption. Moreover, the small bolometric coefficient (1/R ×dR/dT) of graphene (<1%/K) [139, 140] limits the sensitivity of photodetectors based on the readout of the electrical resistance and hinders the use of the bolometric scheme for room temperature applications.

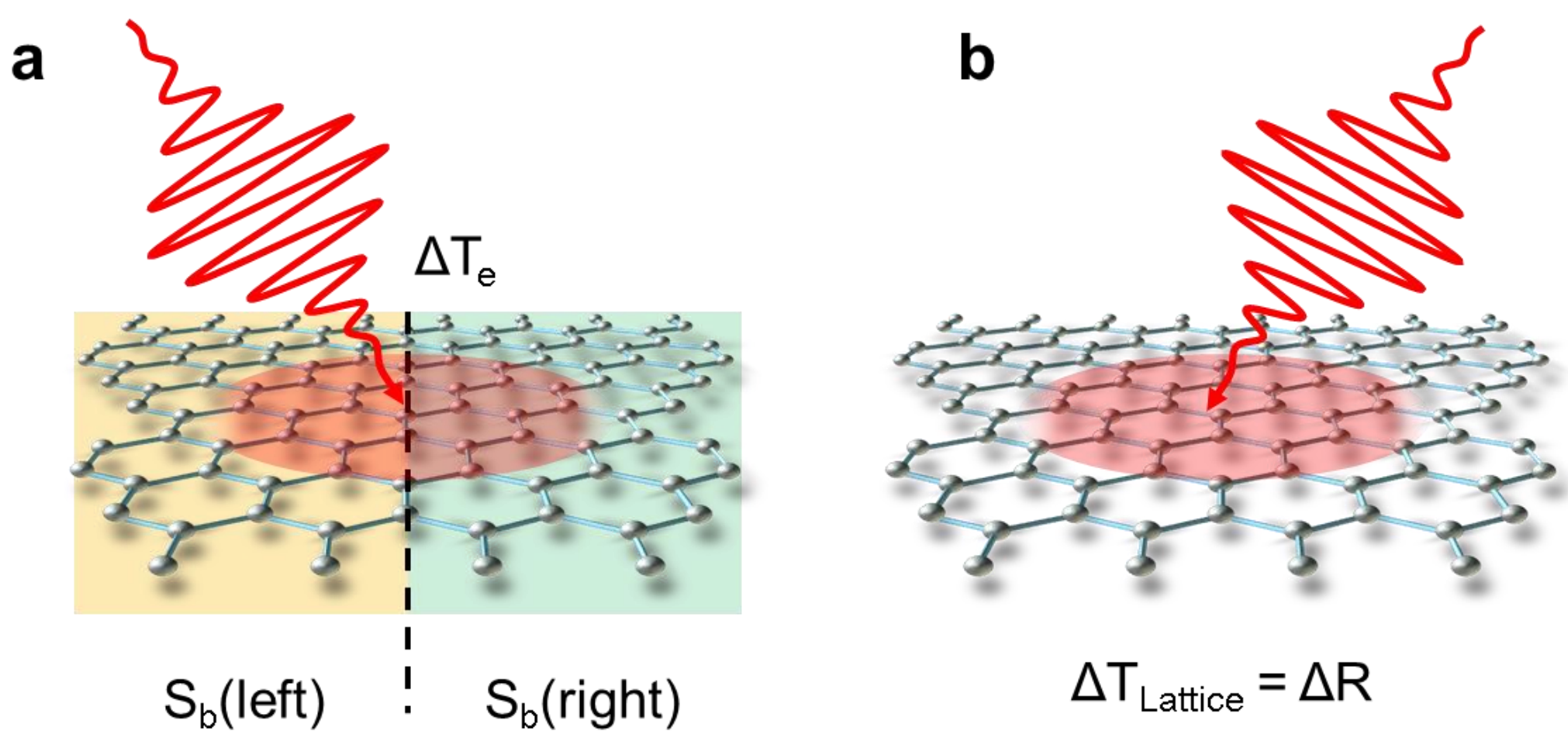

**Figure 2. THz-induced thermal photoresponse mechanisms in graphene. (a)** Photo-thermoelectric (PTE) effect: when the temperature gradient in the electronic distribution (red shaded area) overlaps with a gradient in $S_b$ (the dashed line indicates the junction between two regions with different Seebeck coefficient), a PTE electrical signal is generated. **(b)** The photo-bolometric (PB) effect doesn't need an asymmetry in the sample structure, it suffices the presence of a uniform THz-induced lattice heating, which entails a change in the electrical resistance of the device. A bias voltage is typically needed to obtain an electrical signal.

Another important detection mechanism that can take place in GFETs is the plasma-wave effect [24, 114, 135]. This has been predicted in the mid-1990s by the theoretical works of M. Dyakonov and M. Shur [141, 142] and is therefore also referred to as the *Dyakonov-Shur* effect [25, 143, 144]. Unlike PTE and PB mechanisms, this is not a thermal effect, but relies on the THz-induced plasmonic excitation of the carrier density in the channel of a field effect transistor (FET). The asymmetric coupling of the THz field to the source and gate electrodes produces the simultaneous modulation of carrier density and drift velocity inside the channel, enabling rectification and inducing a net photovoltage at the drain electrode. When plasma waves are overdamped inside the graphene channel, this effect coincides with a general property of FETs, known as *resistive self-mixing* [145, 146], which has been widely demonstrated in sub-THz and THz MOSFETs detectors [147, 148]. RT THz detectors based on antenna-coupled graphene FETs and exploiting the *Dyakonov-Shur* mechanism have also been demonstrated [24, 135], typically in combination with PTE detection [114]. Importantly, at RT, plasma waves are typically overdamped [149], leading to a broadband photoresponse.

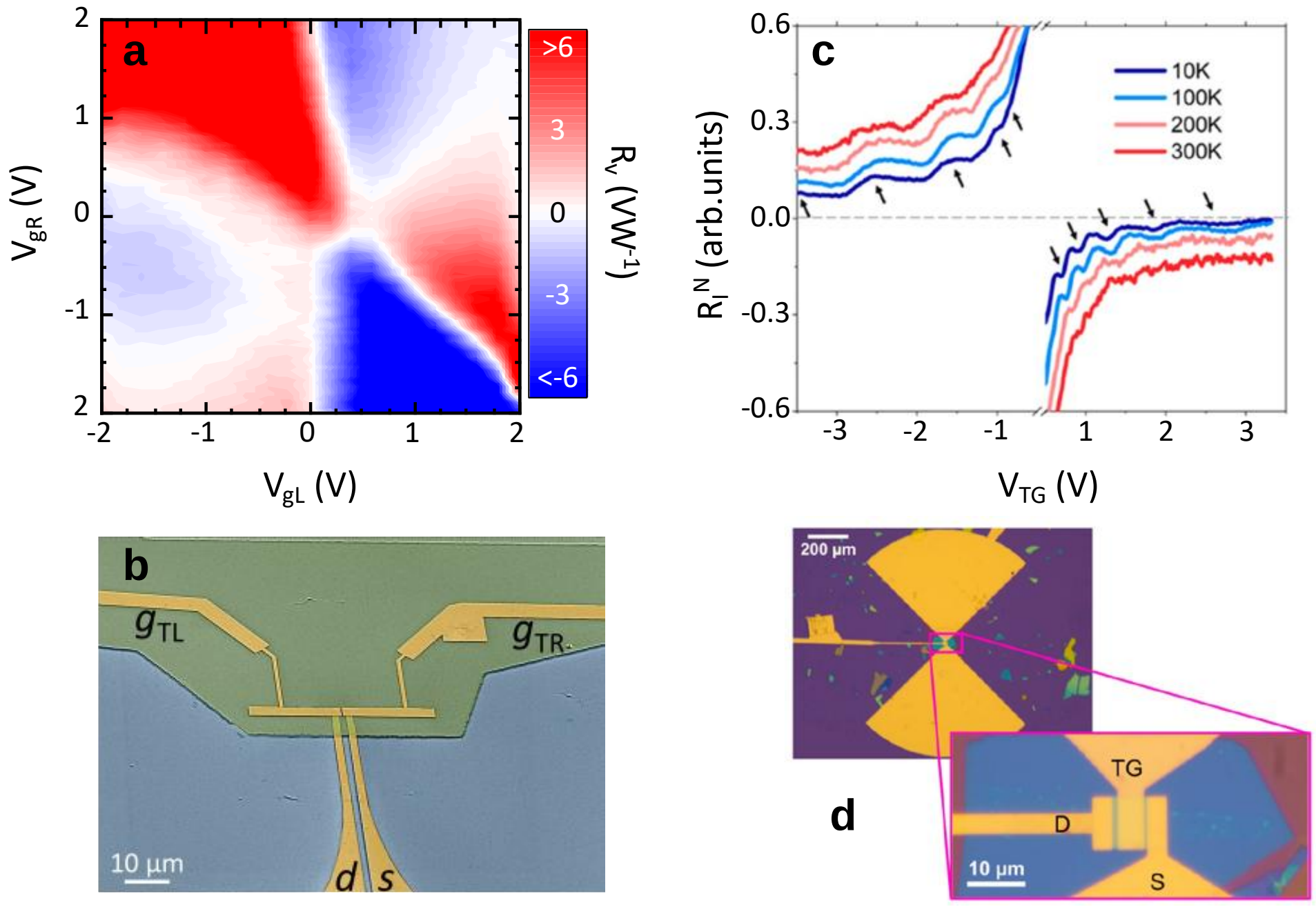

**Figure 3. Physical mechanisms underpinning photodetection. (a)** In a gate-voltage controlled *p-n* junction, the responsivity map, as a function of the gate biases, presents a typical *six-fold* pattern when photodetection is dominated by the PTE mechanism. This pattern stems from the non-monotonic dependence of $S_b$ as a function of gate voltage in a GFET channel. Adapted from Ref. [27]: M. Asgari, et al., ACS Nano, Vol. 15, 17966-17976, 2021; licensed under a Creative Commons Attribution (CC BY) license. **(b)** False-color SEM image of the linear-dipole antenna coupled to the graphene p-n junction. Adapted from Ref. [27]: M. Asgari, et al., ACS Nano, Vol. 15, 17966-17976, 2021; licensed under a Creative Commons Attribution (CC BY) license. **(c)** Zero-bias normalized photocurrent as a function of the top gate voltage at four selected temperatures from 10 K up to 300 K in a high-mobility, top-gated GFET. The presence of oscillations indicates the resonant excitation of plasmonic modes. Adapted from Ref. [150]: J. M. Caridad et al., Nano Letters, Vol. 24, 935-942, 2024; licensed under a Creative Commons Attribution (CC BY) license. **(d)** Optical images of the high-quality GFET THz detector with a bow-tie antenna coupled between source and top-gate electrodes. Adapted from Ref. [150]: J. M. Caridad et al., Nano Letters, Vol. 24, 935-942, 2024; licensed under a Creative Commons Attribution (CC BY) license.

However, by using high quality hBN-encapsulated SLG with carrier mobility $\mu > 50000$ cm$^2$V$^{-1}$s$^{-1}$, RT plasmonic-assisted resonant THz detection has been demonstrated [150], where the excited plasma waves are weakly damped. Furthermore, Bandurin et al. [151] demonstrated that the channel of a FET based on high-quality hBN-encapsulated bilayer graphene (BLG) can act like a resonant cavity for plasmonic waves at low temperature (T = 10 K), opening the interesting possibility to exploit plasmonic interferometry to devise intensity-, frequency- and polarization-sensitive photodetectors [152].

#### 2.2 Quantum Photodetectors

The development of photodetectors with high quantum efficiency, in the far-infrared, has been typically hampered by the low (< 10 meV) involved photon energy, which makes it difficult for a single photon to generate a signal above the noise level of a photodetector. However, the increasing progress of quantum information and the major demand of integrated photonic quantum platforms motivated the engineering of sources and photodetectors with photon counting capabilities even in the challenging THz frequency range. The performance of cryogenic radiation detectors, based on superconducting hot electron bolometers (HEB), have evolved in the last half-century reaching the astrophysical photon noise limit (NEP~$10^{-20}$ $WHz^{-1/2}$) [100]. This kind of receivers has been successfully employed for applications in astronomy [153-155] and spectroscopy [156-158]. However, so far, there are only two reports of efficient architectures for single-photon THz detection: semiconductor quantum dots capacitively coupled to single electron transistors (T = 50 mK, needs magnetic field) [159] and superconductor-based quantum capacitance detectors (T = 15 mK) [155], where THz radiation antenna-coupled to a superconducting absorber generates quasi-particles whose density is measured using a single Cooper-pair box.

In this context, the rapid progresses in graphene-based THz detectors promise ground-breaking advancements towards photon-counting capabilities. A notable example is represented by the recent demonstration of NEP down to 30 $zWHz^{-1/2}$ with a thermal time constant of 500 ns in a bolometer realized in a superconductor–graphene–superconductor (SGS) junction [118] (Figure 4a). An interesting material platform, currently under wide investigation, is bilayer graphene. In general, the addition of a graphene layer modifies the band structure. When the relative twist angle between the two layers is set at 1.1°, the so-called *magic* angle, a moiré pattern gives rise to a long-wavelength periodic potential [160], where flat bands with ultrahigh density of states are formed, and interactions give rise to correlated insulating and dome-shaped superconducting phases with a critical temperature > 3 K. These superconducting magic-angle twisted bilayer graphene (MATBG) devices have been employed as sensitive THz nano-calorimeters [160].

Furthermore, even the inversion symmetric AB-stacked bilayer graphene (BLG) displays interesting features. In its pristine form, it is a zero-bandgap semiconductor.

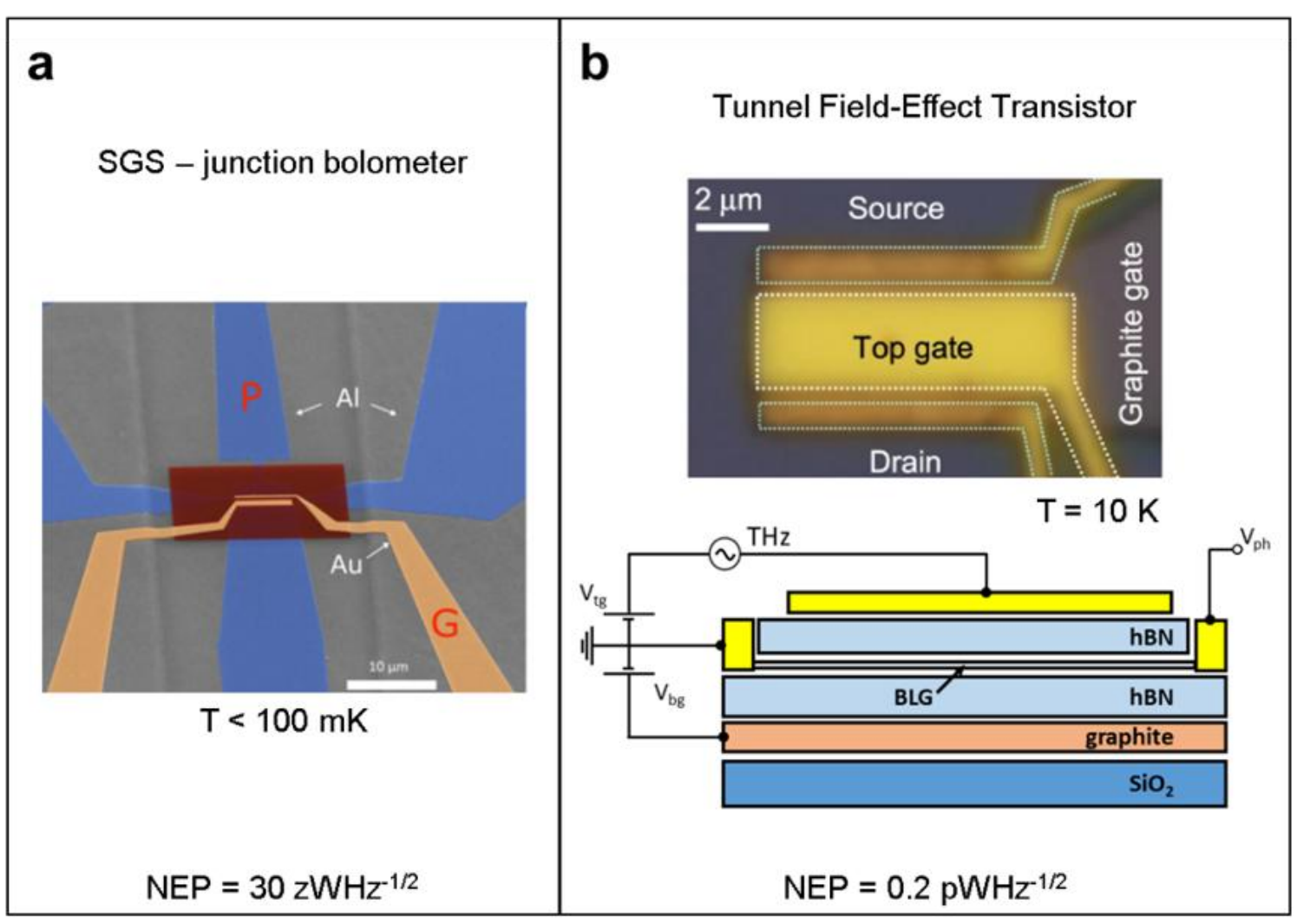

**Figure 4. Cryogenic quantum photodetectors. (a)** A superconductor-graphene-superconductor (SGS) junction (false-color scanning electron microscope image) demonstrated as an ultra-low NEP THz bolometer. The heater and probe signals couple through the aluminum electrodes (port P) to the SGS junction. Adapted from ref. [118]: reproduced with permission from Nature, 586, 47-51 (2020), Copyright 2020, the Author(s), under exclusive licence to Springer Nature

Limited. **(b)** Optical image of a bi-layer graphene-based tunnel field-effect transistor (TFET). The lower inset shows the schematic of the active element. The THz field is coupled between the source and top-gate electrode. The dual-gate architecture provides tunability of the BLG bandgap by the out-of-plane electric field. Adapted from ref. [120]: I. Gayduchenko et al., Nature Communications, Vol. 12, 543, 2014; licensed under a Creative Commons Attribution (CC BY) license.

However, if the two individual graphene layers are made inequivalent by breaking the out-of-plane inversion symmetry, an energy gap between low-energy bands opens at the former Dirac crossing point [161-163]. Interestingly, this entails the possibility of tuning the bandgap of BLG by applying an external electric field along the out-of-plane direction, e.g. by employing a double (top-bottom) gates configuration [162], up to 120 meV [164]. Such a system is intrinsically endowed with *striking* functionalities: (i) the conductance of a graphene bilayer channel could be switched over a wide range at a speed which is limited by gate-voltage switching, (ii) the band alignment along a channel can be tailored by using multiple gated areas, (iii) the optical properties can be modified on demand. These features have been exploited to different extents in many recent optoelectronic devices operating across the infrared spectrum and down to the THz domain. For example, tunnel field-effect transistors (TFETs) based on dual gated BLG coupled to broadband THz antennas have been demonstrated as sensitive sub-THz detectors [120-], with NEP ~200 fW/√Hz (Figure 4b). More recently, large-angle twisted double bilayer graphene (TDBG) have been used as a narrow-gap semiconductor (~10 meV, induced by the crystal-field of large-angle-stacked BLG layers [163]) for sensitive and ultra-broadband detection spanning the frequency range from 3 THz to 150 THz in a single device [110].

## 3. Amplitude, polarization and phase modulators with graphene

Optical modulation plays a crucial role in infrared photonic and optoelectronic applications, including optical interconnect, environmental monitoring, biosensing, medicine and security applications. With the rapid growth of mobile data traffic [165, 166] and the demand for unallocated frequency ranges for next generation wireless communications (6G) [167], different approaches for THz frequency modulation were demonstrated, based on III-V semiconductors such as silicon and gallium arsenide, or employing 2DEG in AlGaAs/InGaAs heterostructures, with modulation speeds up to 14 GHz and room temperature operation [168, 169]. Beside these technologies, graphene has been an interesting functional material of choice for devising THz modulators, mainly due to the fact that the material optical response, in the THz range, is dominated by free carrier (Drude) intraband absorption, which is proportional to the real part of the complex conductivity $\sigma$ [50].

From equation (e1), it originates that $\sigma$ depends on the concentration ($n$) and temperature ($T_e$) of graphene carriers. Therefore, absorption can be varied by actively controlling $n$ and/or $T_e$. This principle is at the base of graphene-based modulators, which typically take advantage of its wide carrier concentration tuning range (up to ~$10^{14}$ $cm^{-2}$) [170]. Moreover, strong and broadband absorption in the far-infrared, fast carrier dynamics and ease of integration with standard photonic platforms [171] such as waveguides, micro-resonators and fibers, can potentially enable disruptive large-scale applications for graphene. Hence, active phase [172, 173], frequency [174], amplitude [169, 175, 176] and polarization [177, 178] modulators realized in graphene have been reported in the THz frequency range.

In this section, we broadly describe the different approaches that have been proposed, highlighting the advantages or dis-advantages of graphene-based modulators with respect to other material platforms. In general, different figures of merit are used to describe the performance of an optical modulator [179]: the modulation depth (MD) for amplitude modulators [109], the optical activity (OA) or circular dichroism (CD) for polarization modulators [179] and phase shift (PS) for phase modulators [173]. Beside these, a key parameter is the modulation (or re-configuration) speed, which describes how fast a bit of information could be encoded in the modulated signal and is typically evaluated as the cut-off frequency, at which the reference figure of merit is decreased by 3dB.

### 3.1 Amplitude modulators

Optical amplitude modulation is a fundamental cornerstone of optoelectronic circuits and is typically achieved by changing the optical absorption of the employed materials, primarily semiconductors. An ideal intensity modulator is fast, broadband, and deep (MD close to 100%), with a high pristine transparency. In the THz frequency range, amplitude modulation can be achieved by exploiting three main system configurations: integrated all-optical modulators, all-electronic modulators or microelectromechanical systems. A detailed overview of these schemes can be found in recent topical reviews [180, 181]. In the case of graphene, the most technologically relevant means of controlling the absorption and achieving active modulation are optical pumping [182, 183] and the electrostatic gating of the carrier density and carrier temperature [109].

THz all-optical modulation is based on the fast generation of free-carriers by a *pump* beam, with photon energies typically larger than the bandgap ($E_g$) of the employed material ($\hbar\omega > E_g$). The photo-generated carrier density produces a temporary increase of the conductivity in the modulator, whose response to the incoming (*probe*) THz radiation can thus be actively controlled. This strategy is broadband and becomes progressively more effective at lower frequencies, due to free-carrier absorption. It is also highly versatile, allowing, for example, to control the spatial distribution of free carriers in a substrate that, in turn, can act as a programmable spatial THz light modulator [184, 185]. Optical modulation has historically been a convenient solution for THz amplitude modulators, as it allows for the use of silicon [186] or III-V semiconductors [187], such as germanium or GaAs, as active substrates along with near-infrared pumping light.

THz modulators based on GaAs and pure silicon have been explored [188, 189], suggesting that amorphous Si and Au-doped Si (Au:Si) could be suitable for achieving fast switching times. The speed limit of these modulators is determined by the recombination time ($\tau_r$) of photo-generated carriers, with $\tau_r > 1$ ms for silicon [190] and $\tau_r < 1$ ps for low-temperature grown GaAs [191]. Although GaAs has the shortest lifetime and may therefore offer faster modulation, it also has higher absorption [192] and a higher, more dispersive refractive index [193] compared to crystalline Si. For this reason, Si is considered the material of choice for the development of all-optical modulation schemes in the far-infrared. Recently, broadband (0.5-7 THz) all-optical high-speed THz Si-based modulators have been realized in an optical fiber with a 210 µm diameter gold-doped silicon core, achieving a 96% MD [194]. It is important to note that MD values typically decrease when transitioning from integrated solutions (waveguides, fibers, etc.) to free-space modulators, with typical values ~70% in transmission [185]. In this context, combining graphene with semiconductor substrates can enhance the performance of all-optical modulators. For instance, the integration of monolayer graphene with Si has resulted in MD up to 80% (Figure 3a) [182], and when combined with Ge, MD can increase from 64% (substrate alone) up to 94% by pumping at 1550 nm [183]. This improvement is attributed to the transfer of photo-generated carriers from the substrate to the graphene layer, where they experience a higher mobility, and in turn, a higher conductivity.

Importantly, the integration of graphene allows modifying the optical response of a system by electrostatically tuning its carrier density [109]. This concept has driven the development of all-electronic modulators. In contrast to the all-optical scheme, the electrical approach simplifies the modulation system, removing the need for external optics and additional sources. Indeed, all-electronic modulation in graphene is typically obtained by simply changing an applied electrostatic gate voltage [195]. Since the pioneering works on large-area graphene by Maeng and Sensale-Rodriguez in 2012 [175, 196] (Figure 5a), all-electrical THz modulators based on large-area graphene have been developed by many research groups (Figure 5b,c) [29, 30, 197], leading to impressive results in terms of performances: MDs beyond 60% and modulation speed in the 100 kHz range have been achieved. Recent topical reviews have thoroughly addressed the evolution of these advancements, marking the major challenges and the possible weaknesses of the many different approaches [169, 179].

The two main hurdles that are faced using graphene (and 2D materials in general) for electrically driven modulation of THz radiation are (i) the low absorption cross-section of atomically thin layers and (ii) the typically low reconfiguration speed of large-area graphene by a global gate electrode. The most promising strategy to simultaneously address these two issues is represented by integrating graphene into metamaterial (MM) matrices: planar artificial structures composed of many unit cells or metallic structures, whose size is typically smaller than the radiation wavelength. With a relative ease of fabrication, high-efficiency, and spectral tuning, metasurfaces [198] can lead to "designed" responses to electromagnetic fields, offering unprecedented functionalities for beam shaping, polarization control and wavefront generation. In combination with graphene, a metamaterial can act as *enhancer* for light-matter interaction, simultaneously lowering the interaction area, thereby enabling a faster reconfiguration.

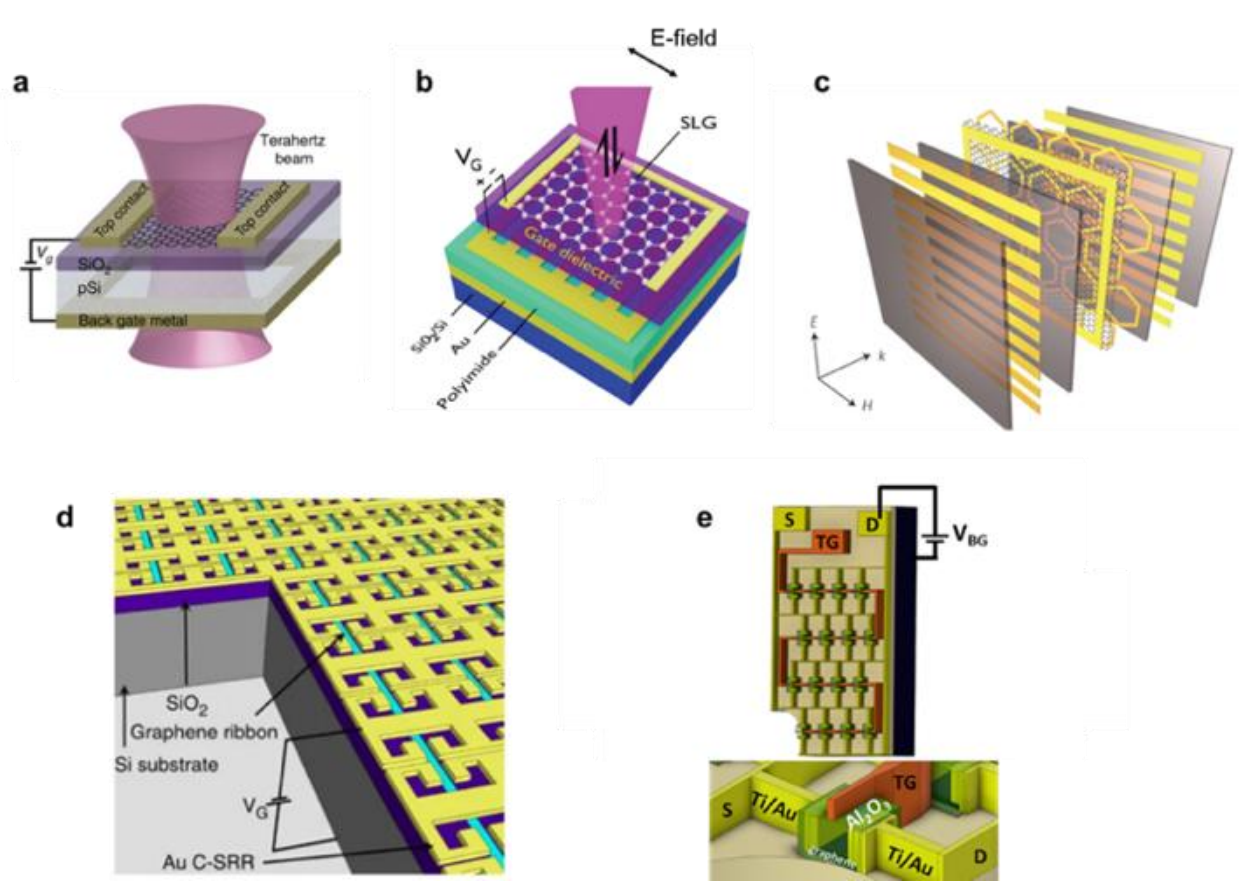

**Figure 5. Graphene-based amplitude modulators: architecture evolution towards high cut-off frequency devices. (a)** Schematics of a broadband tunable modulator based on large-area graphene operating in transmission. Adapted from ref. [175]: reprinted with permission from Nano Lett. 12, 4518–4522. Copyright 2012 American Chemical Society. **(b)** A grating-gated reflection modulator based on a single-layer graphene (SLG) in a Salisbury-mirror geometry. Adapted from ref. [29]: A. Di Gaspare et al., Adv. Funct. Mater. 31, 2008039, 2021; licensed under a Creative Commons Attribution (CC BY) license. **(c)** Gate controlled active graphene metamaterial, composed of a SLG deposited on a layer

of hexagonal metallic meta-atoms. Adapted from ref. [195]: reprinted with permission from Nature Materials 11, 936–941. Copyright 2012 Springer Nature Limited. **(d)** Three-dimensional schematics of a tunable metasurface composed of graphene ribbons and electric-field-coupled complementary split-ring resonators (C-SRRs). Adapted from ref. [199]: P. Q. Liu et al., Nature Communications 6, 8969 (2015); licensed under a Creative Commons Attribution (CC BY) license. **(e)** Resonant metamaterial based on graphene-coupled meta-atoms. Adapted from ref. [202]: A. M. Zaman et al., Nanotechnol. 5, 1057422 (2023); licensed under a Creative Commons Attribution (CC BY) license. The progressive reduction of the light-matter interaction area reduced the overall electrical inertia of the system, leading to progressively high modulation speeds > 3 GHz in (e).

Importantly, tuning the conductivity of MM-integrated graphene entails both a change in the overall absorption and a change in the frequency response of the metasurface itself, resulting in improved MD performance. Modulators based on the MM-graphene combination have been developed in the last few years [199, 200] (Figure 5d) reaching amplitude MD >80% and reconfiguration speed >3 GHz (Fig. 5e) [201, 202]. These results hold concrete prospects for applications in THz communications and real-time imaging.

It is worth mentioning that the MM concept can be easily applicable to a plethora of different material system for devising electrically driven modulators. As notable examples electrically tunable two-dimensional electron gases (2DEGs) have been realized with semiconductor III-V (GaAs [203], AlGaN/GaN [204], InAlN/AlN/GaN/AlN/GaN [205]) heterostructures, where MD up to 93% have been achieved, with -3 dB cut-off frequency of ~1 GHz. More recently, sub-THz (300 GHz) large-scale programmable metasurfaces have been realised using arrays of complementary metal–oxide–semiconductor (CMOS)-based chip tiles, achieving GHz modulation speed [206-, 207]. These and other promising material platforms are described in a recent topical review [179].

Another possible strategy to maximize the optical coupling between 2D-graphene and a THz beam is represented by the direct integration of graphene with semiconductor-based THz sources, e.g. quantum cascade lasers (QCLs) [208, 209]. This approach is made possible thanks to the extraordinary fabrication versatility and mechanical resilience of graphene, that enables its deterministic transfer on virtually any host substrate/architecture. By tuning the strong intraband absorption, the emission of the hybrid QCL-graphene architectures can be tailored both in intensity and spectral content, with operating speed in the 100 MHz range [208]. When coupled to QCLs in an external-cavity configuration, graphene-based modulators can modify and stabilize the emission of the laser by controlling the optical feedback. This approach has been used to increase the operational dynamic range of QCL frequency combs (FCs), opening interesting prospects for short pulse generation, phase-locking, frequency tuning/chirping, and metrological referencing [29, 30].

### 3.2 Polarization modulators

The development of THz polarization modulators has recently drawn the attention of researchers thanks to their potential use in THz free-space communications, where protocols such as polarization shift keying [210] and polarization division multiplexing [211] are emerging as potential bandwidth-savers. Moreover, an active control of THz light polarization enables the probing and characterization of inherently chiral biomolecules, such as DNA, RNA, and proteins [212]. In this regard, the combination of graphene with metasurfaces offers a versatile tool to address the objectives of strong rotation of the polarization plane (large optical activity) and speed, enabling the realization of artificial chirality and multi-layer architectures.

In the THz frequency range, polarization modulators are primarily realized by using metasurfaces, exploiting either micro-electro-mechanical systems (MEMS) or tunable meta-atoms with electrically-driven adaptive chiral response. In this latter case, different material platforms have been used: $VO_2$ grid polarizers [213], liquid crystals [214] and graphene [215, 216, 217] (Figures 6a,b,c). Polarization conversion efficiency up to 95% has been achieved in gated double-layers of graphene. More recently, polarization modulation with dynamic conversions from linear to elliptical polarization states was reported in graphene-loaded metamaterials, achieving an almost independent control of circular dichroism (CD) and OA. Importantly, this has been achieved in an all-electronic system with reconfiguration speed > 1 GHz [202].

### 3.3 Phase Modulators

Devising direct phase modulators in the THz range is very challenging. This is primarily due to the lack of a compelling physical mechanism to drive this effect: broadband free-carrier absorption typically has a minor effect on the phase of the output THz radiation. Moreover, the active manipulation of the phase, typically entails an amplitude variation as a collateral effect. Nevertheless, controlling the phase of a THz wave opens new perspective in this frequency range, encompassing beam steering [218, 219], polarization conversion [220], generation of vortex beams [221], digital encoding for free-space [222] and guided [223] THz communications. Different approaches have been recently implemented for the development of electrically driven THz phase shifters, typically based on MEMS [224-] and electrically tunable materials, such as liquid crystals [225, 226], graphene [227, 228] or vanadium dioxide [229, 230], which enable real-time controllable phase gradients or phase distributions. An in-depth picture of the available technologies can be found in recent topical reviews [231, 232]; here, we focus on recent advancements on free-space graphene-based active phase-shifters.

Large-area graphene has been used to realize a tunable impedance surface to modulate both the amplitude and phase of a THz beam [228] (Figures 6d,e). The idea of tuning the phase of a reflected wave by tuning the graphene complex conductivity has been exploited to realize Brewster-angle wideband phase modulators [172] and to implement phase rotation in resonator-based MM architectures. In 2018, THz beam steering by using a graphene-based reconfigurable reflectarray (Figures 6f,g) have been demonstrated. The electrically-controllable unit cells composing the metasurface consist of a THz bow-tie antenna that funnels the impinging radiation on a gate-tunable sub-wavelength graphene active element. By adjusting the gate voltages on individual segments (columns) of the reflectarray, the deflection angle was adjusted, spanning a steering range of ~25°. Importantly, large phase-shifts can be obtained in graphene-loaded resonators. In particular, by acting as an electrically-tunable lumped element, graphene can introduce a spectral detuning of the resonator response. This concept has been used by Miao et al. [227] to realize phase rotation of ±180°.

Importantly, controlling the phase degree of freedom of a beam with a metasurface allows for achieving focusing and coupling linearly polarized light into vortex beams with different topological charge [234]. Being efficient phase-shifters, graphene coupled metamaterials can be utilized to generate THz beams with well-defined optical angular momentum (OAM) [235-237], thus opening important perspectives for information multiplexing in optical communications. A comprehensive description of the various metasurface technologies that can be utilized for this purpose can be found in a recent topical review (ref. [235]).

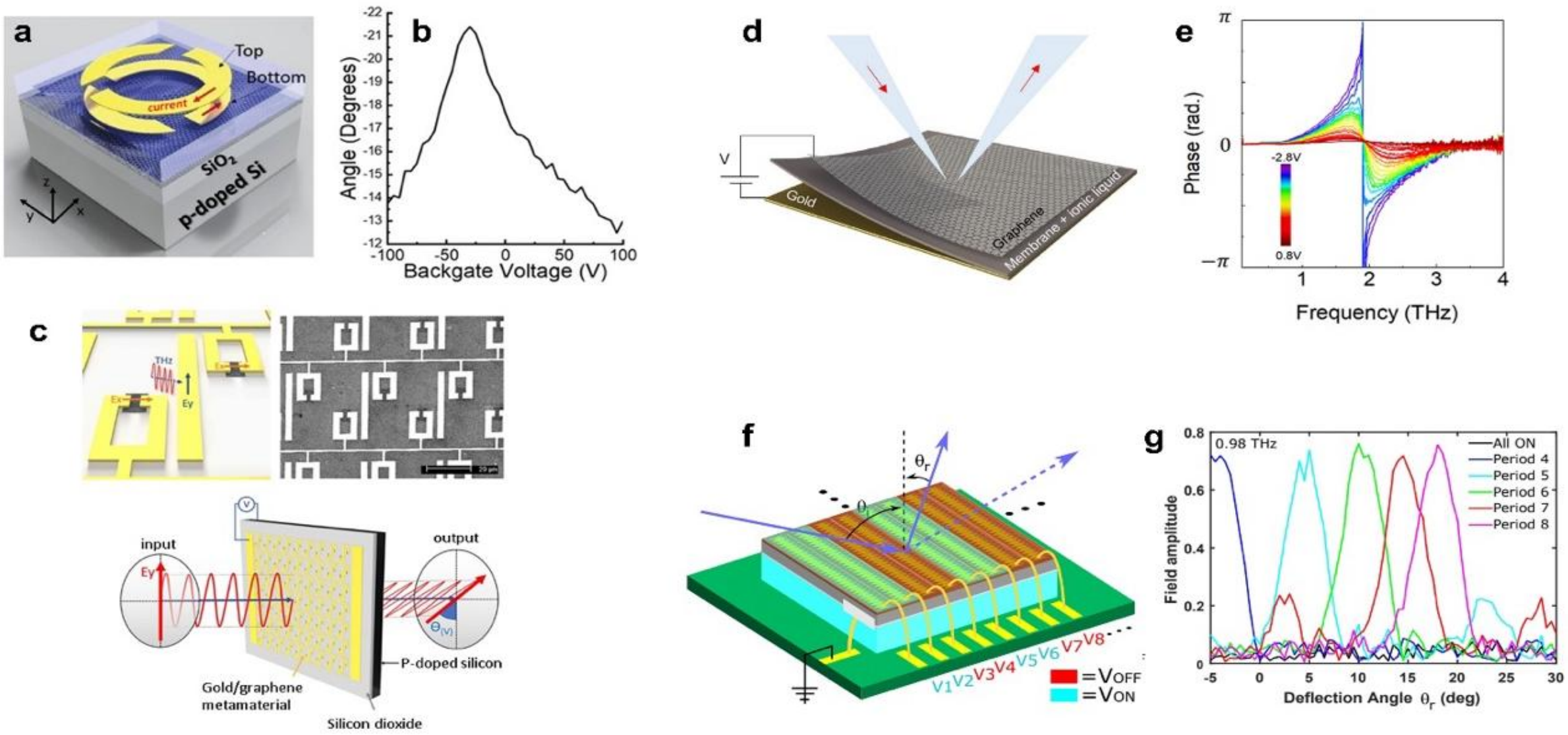

**Figure 6. Polarization and phase modulators based on graphene. (a)** Example of a chiral metamaterial for polarization rotation: the unit cell of the multi-layer structure is based on conjugated rings. The two currents excited in the upper and lower rings are antiparallel and give rise to a magnetic dipole and magnetic field component along the *y*-direction, responsible for the polarization rotation. Adapted from Ref. [216]: S. J. Kindness et al., Adv. Optical Mater. 8, 2000581 (2020); licensed under a Creative Commons Attribution (CC BY) license. **(b)** Rotation of the polarization plane measured at 1.94 THz frequency as a function of the gate voltage. Adapted from Ref. [216]: S. J. Kindness et al., Adv. Optical Mater. 8, 2000581 (2020); licensed under a Creative Commons Attribution (CC BY) license. **(c)** Electrically-driven linear polarization rotator. Schematic of device unit cell and SEM image showing a section of the device array. Graphene patches are placed in the capacitive gap of the resonators, the area corresponding to the higher field-enhancement. Lower inset: illustration of the working principle of a polarization rotation device. Adapted from Ref. [215]: S. J. Kindness et al., ACS Photonics 6, 1547–1555 (2019); licensed under a Creative Commons Attribution (CC BY) license. **(d)** A graphene-based phase modulator consisting of a large-area single layer graphene on a porous thin (20 μm) membrane of polyethylene with a gold reflector at the back. **(e)** Phase variation as a function of the impinging radiation frequency for different applied back-gate voltage. Adapted from Ref. [228]: N. Kakenov et al., 2D Mater. 5, 035018 (2018); licensed under a Creative Commons Attribution (CC BY) license. **(f)** Schematics of the large-area programmable reflectarray used for beam steering. Beam deflection can be achieved by using periodic or quasi-periodic patterns of *on* and *off* columns. **(g)** Beam profiles obtained with different control voltage patterns. Adapted from Ref. [233]: licensed under a Creative Commons Attribution (CC BY) license.

## 4. Saturable Absorbers

Saturable absorption is a third-order non-linear optical process that induces, as an effect of field-dependent nonlinear optical conductivity, an intensity-dependent change in the refractive index of a material. Saturable absorption can be in principle activated for both intraband and interband transitions. Owing to its fast (hundred femtosecond) carrier dynamics [69, 63], the large (~10 %) absorption of incident light at terahertz frequencies [238], and the possibility to saturate this absorption with relatively low incident radiation, graphene is a potential saturable absorber material [239], at THz frequencies. In this range of frequencies, the optical conductivity of graphene is mainly determined by intraband transitions (with a less dominant interband picosecond timescale contribution due to optical phonon cooling [122]), meaning that its optical absorption can be easily modulated by electrically or optically controlling its Fermi level [240, 241]. Combining graphene with metamaterial patterns provided to be a valuable trategy to induce local modulations of refractive index via optical Kerr effect [240].

The optical non-linearity of graphene is a result of its carrier thermodynamics. Under an impinging THz external electric field, electrons are heated, and the excess energy re-distributed by electron-electron scattering on 10-100 fs time scales; as a result, the free-carrier population share an elevated electron temperature [50]. The same timescales govern the electron transport lifetime, determined by momentum scattering events. Given the relatively slow (~ps) oscillation of THz waves, before thermalizing, electrons experience several scattering events within one period, resulting in a diffusive collective motion. This leads to a *quasi*-instantaneous increase in electronic temperature and a concomitant decrease in conductivity. This scenario corresponds to saturable absorption: the stronger is the field, the smaller the conductivity becomes [72, 242-244].

The described physical mechanism entails a very efficient nonlinearity, enabling saturation with relatively weak fields of ~10 $kVcm^{-1}$, and requiring peak powers $10^6$ times smaller than those needed to achieve optical or IR nonlinearity by means of coherent Bloch oscillations, as reported experimentally [73]. At THz frequencies, graphene exhibits possibly the highest nonlinear coefficients of all known materials to date [73].

The first demonstration of THz nonlinearity in graphene was achieved through nonlinear THz-TDS and THz pump–THz probe spectroscopy [244, 245]. By exposing CVD-grown graphene to intense single-cycle THz pulses with a maximum fluence of 190 $\mu Jcm^{-2}$, which corresponds to a peak electric field strength of ≈100 $kVcm^{-1}$, THz-field-induced transparency was observed. The enhancement in nonlinear transmission was attributed to saturable absorption effects, which are linked to a decrease in the intraband conductivity of the doped graphene sample, with a saturation fluence of 20 $\mu Jcm^{-2}$.

Graphene saturable absorbers (GSAs) have been utilized to demonstrate mode-locked lasers at frequencies ranging from the visible to the IR [246-251]. The first report of THz saturable absorption using multi-layer graphene grown on the carbon-face of silicon carbide achieved a maximum modulation depth of approximately 10% [252] (Fig. 7a-d). Saturable absorption was also demonstrated in graphene grown via CVD on Nickel, exhibiting a comparable intensity modulation [253] (Fig. 7e).

THz saturable absorption was subsequently achieved in samples realized through transfer coating and inkjet printing of single and few-layer graphene films prepared by liquid phase exfoliation of graphite [67]. Open-aperture *z*-scan experiments with a 3.5 THz QCL showed a transparency modulation ∼80% (Fig. 7f), which is almost one order of magnitude larger than that reported for other materials at THz frequencies [254].

Following these initial examples of graphene-based THz SAs, more complex and integrated configurations have been developed. For instance, a 15-layer graphene sample was transferred onto a 0.3 mm thick silicon (Si) mirror and integrated with a THz QCL facet. This created a composed cavity that helped self-stabilize the intermodal beat note linewidths of the laser [255]. Another sophisticated architecture of a THz saturable absorber mirror based on SLG involved a ≈12.5 µm thick polypropylene spacer, which defined a cavity filled with liquid electrolyte between the SLG and a 300 nm thick Au layer. The advantage of electrolyte gating (EG), in a supercapacitor geometry [249], is the large (a factor ~10) $E_F$ tuning at low bias (≈1 V). This is possible due to the self-forming ultrathin electrical double layer at the graphene-electrolyte and Au-electrolyte interfaces, which generates large electric fields (≈$10^9$ V $m^{-1}$) at nanometer scales without electrical breakdown. This configuration has great potential for the development of graphene optoelectronic components [249], although it has limited modulation speeds. In ref. [30], electrically tunable THz saturable absorption was demonstrated with a large area (≈10×10 $mm^2$) epitaxial SLG, transferred onto a quartz substrate (500 μm thick). With this scheme, a reflectivity modulation of 60% was achieved, with a saturation intensity of 4.5 $Wcm^{-2}$ (Fig. 7g-h).

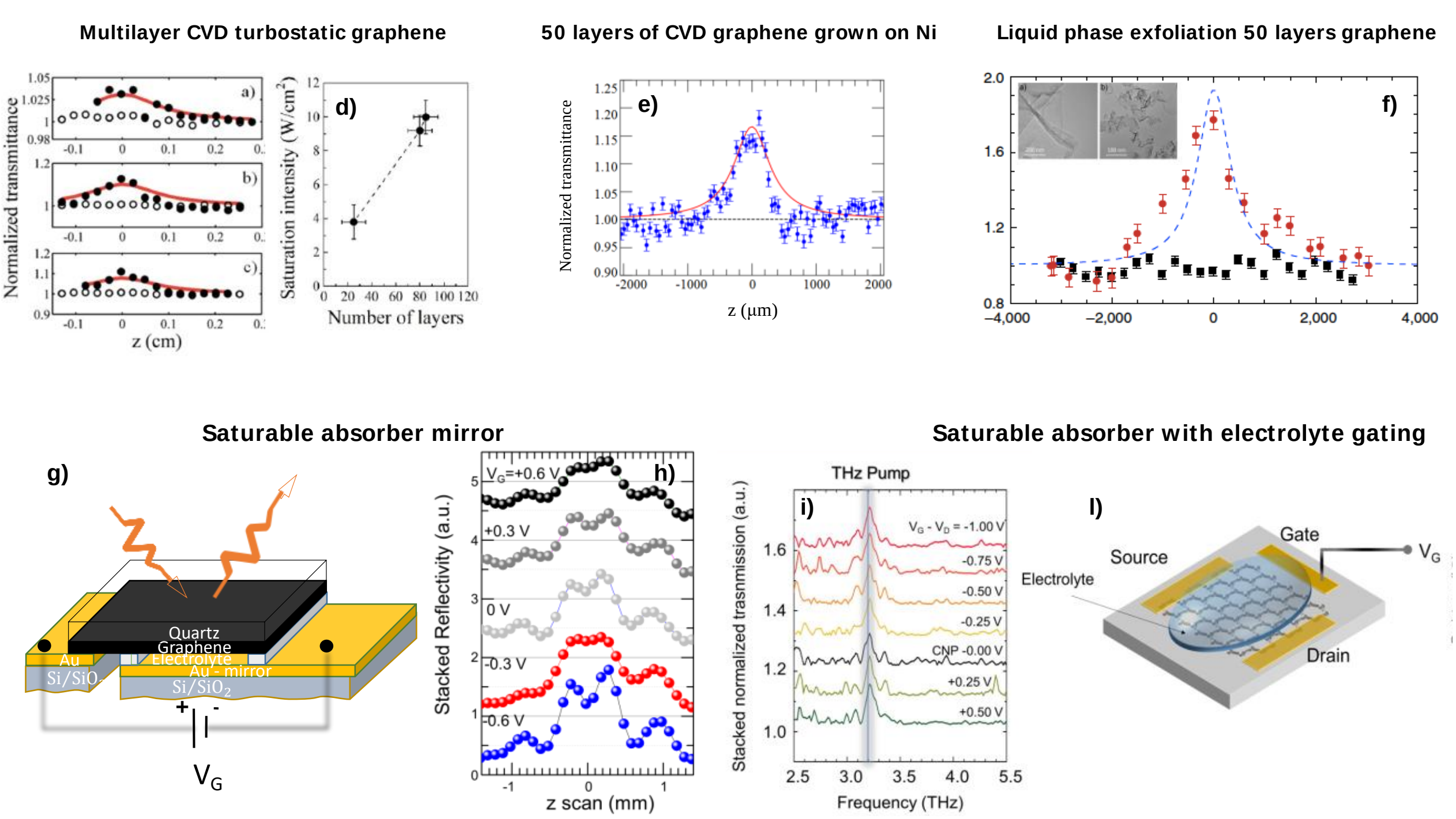

**Figure 7. Terahertz saturable absorbers. (a-c)** Open aperture z-scan traces measured on a turbostratic multilayer graphene sample grown on a substrate of silicon carbide, with $n$ = 25 (a), $n$= 80 (b) and $n$ = 85 (c) layers (solid dots), and of the substrate (empty dots). The fit curves assuming the simple two-level saturable absorber model are shown as red lines. **(d)** Plot of saturation intensity as a function of the number of layers. From Ref. [252]: F. Bianco et al., Opt. Express 23, 11632–11640 (2015); licensed under a Creative Commons Attribution (CC BY) license. **(e)** z-scan normalized transmittance of a multilayer (n = 50) graphene sample grown by CVD on Nickel, and probed with a THz QCL, compared with the z-scan normalized transmittance traces of the bare $Si/SiO_2$ substrate (dashed horizontal line). The red line represents the fit using the following equation for the transmittance: $T(z) = \left[1 - \alpha_0 + \alpha_S - \frac{\alpha_S\left(1+\frac{z^2}{z_R^2}\right)}{1+\frac{z^2}{z_R^2}+\frac{I_0}{I_S}}\right]\frac{1}{1-\alpha_0}$, where $\alpha_{NS}$ and $\alpha_S$ are the non-saturable and saturable components of the linear absorption, $I_0$ is the beam intensity at the focal point, $z_R$ is the Rayleigh length, $I_S$ is the saturation intensity and $\alpha_0 = \alpha(I = 0) = \alpha_{NS} + \alpha_S$ is the linear absorption. From Ref. [253]: M. S. Vitiello et al., J. Phys. Mater. 3 014008 (2020), licensed under a Creative Commons Attribution (CC BY) license. **(f)** $z$-scan normalized transmittance at 3.5 THz of an inkjet-printed water-based graphene saturable absorber (red dots), compared with the $z$-scan traces of the $Si/SiO_2$ substrate (squares). The error bars correspond to the uncertainty interval on the measured normalized transmittance. The dashed line presents the fit of the saturated transmittance T(z). From Ref. [67]: V. Bianchi et al., Nat. Commun. 8, 15763 (2017), licensed under a Creative Commons Attribution (CC BY) license. **(g)** Schematic diagram illustrating the layout of device saturable absorber mirror under top illumination. It consists of a single layer graphene (SLG) on quartz, an electrolyte layer, and an Au mirror layer. Polypropylene spacers of thickness 12.5 μm define a quarter wavelength λ/4 Salisbury mirror-like cavity. The SLG conductivity can be adjusted using the ionic liquid electrolyte gate (EG) by applying an electrostatic gate voltage $V_G$. The transparent top quartz substrate seals the graphene/EG system, enabling operation in the THz range in reflection mode. **(h)** z-scan stacked reflectivity measured at different gate doping levels, normalized with the corresponding curves acquired on the reference sample. From Ref. [30]: A. Di Gaspare et al., Adv. Optical Mater. 10, 2200819 (2022), licensed under a Creative

Commons Attribution (CC BY) license. **(i)** Stacked transmission spectra measured when the sample is illuminated by a THz QCL, obtained by taking the ratio of the experimental transmissions measured on SLG. These measurements were taken while varying $V_G$ with a step size of $\Delta V_G = 0.25V$. From Ref. [241]. **(l)** Schematic diagram of the device: SLG is deposited on quartz, with source, drain, and gate electrodes. The SLG conductivity is tuned using an ionic liquid electrolyte gate, with $V_G$ applied in steps of $\Delta V = +0.3V$. The transparent top quartz substrate contains the graphene system. From Ref. [241]: A. Di Gaspare et al., ACS Photonics 10, 3171-3180 (2023), licensed under a Creative Commons Attribution (CC BY) license.

A similar approach has been also adopted in transmission [241]. A SLG top-gated field-effect transistor is fabricated on a 0.5 mm-thick quartz substrate, utilizing an ionic liquid top-gate. An electrolyte ionic liquid *N*,*N*-diethyl-*N*-methyl-*N*-(2-methoxyethyl ammonium-bis-trifluoromethanesulfonyl imide) of 99.9% purity exhibits a significant electrochemical power owing to the achievable large surface electric field (∼10–20 MV/cm). This allows for an $E_F$ tunability up to 1.2 eV. The final device operates in transmission mode and has an access optical window of ~$6 \times 6$ mm$^2$. Transmission experiments using a z-scan method reveal an increase in transmission at the focal plane, indicating the presence of SA [257], with a measured transmission modulation 61 ± 6% and saturation intensity 5.63 ± 0.22 W/cm$^2$ [241] (Fig. 7i-l).

Beside saturable absorption, many optically induced nonlinear phenomena have been observed in graphene, including high-harmonic generation (HHG) [258], four-wave mixing [259], and self-phase modulation [260]. These phenomena occur under the influence of single or multi-cycle driving fields with peak electric fields up to MV/cm and peak intensities up to TW/cm$^2$, in the near infrared [258]. Importantly, at low frequencies (≤ 2.15 THz), the nonlinearity of graphene can be driven with much weaker electric field strength, with peak electric fields up to 80 kV/cm [79]. This property is particularly relevant for the generation of THz radiation in graphene by frequency up-conversion, which will be discussed in the following sections.

## 5. Miniaturized THz frequency lasers

The recent progresses in the photonic engineering of semiconductor lasers, combined with the state of the art nanofabrication capabilities and ingenious design concepts for electromagnetic components, have resulted in unprecedented control over the motion of electrons and photons at the nanoscale. For instance, microcavities [261], photonic crystals [262], pseudo-random [263] and random [264] crystals, as well as surface-guided modes including plasmon- and phonon- polaritons [265], have emerged as effective strategy to confine light in small volumes and at pre-defined frequencies. Additionally, the possibility to integrate non-linear materials, as graphene with existing technologies, such as semiconductor heterostructure lasers (QCLs), holds great promise for exploring new non linear phenomena, in such a class of lasers

### 5.1 Mode locking

THz QCLs experienced a remarkable progress in the past decade. In the more conventional scheme, the operating temperature now reaches -12°C [266], the output power is > 2 W [267], the frequency coverage spans from 1.2 THz [268] to 6 THz, and the frequency tunability reached 10% of the emission frequency [269, 270]. QCLs exhibit unmatched compactness and the possibility to be stabilized both in frequency, phase, and amplitude [271, 272], or operate as frequency combs [273]. Additionally, THz QCLs offer quantum-limited spectral purity that surpasses any other semiconductor source [274]. The time is ripe for comprehensive studies on the non-linear dynamics of THz QCL modes, targeting mode-locking operation in a regime in which pulse generation is self-starting and self-maintaining.

Typical strategies for transitioning a laser from continuous-wave (CW) to pulsed operation include gain or *Q*-switching, periodic external modulation of gain and loss (active mode-locking) or the presence of a non-linear element in the laser cavity (saturable absorber) (Fig. 8a), which provides self-modulation of both the real and imaginary parts of the net gain (self or passive mode-locking) [275]. However, applying these mechanisms to QCLs (electrically driven intersubband lasers), is not straightforward. This results from the fact that gain in the semiconductor active medium of QCLs recovers from its saturated value much faster (picosecond) than the cavity round-trip time (∼70 ps for a 3 mm cavity) and photon lifetime. The short gain recovery time is due to strong electron-phonon interactions in the polar gain medium of QCLs, which induce ultrafast non-radiative intersubband relaxations. This phenomenon complicates the generation of stable ultrafast laser pulses using mode-locking.

The mode-locked operation of QCLs has been a highly debated topic, with the initial reports of passive mode-locking of QCLs operating in the mid-infrared [276] later re-interpreted as coherent dynamic instabilities caused by the short gain recovery time of these “class A” lasers [277]. However, in a QCL, owing to the possibility to engineer the electronic wave functions on a nanometer scale, carrier transport and gain spectrum can be intentionally customized.

It was only in 2009 that pulsed operation of a mid-IR QCL was demonstrated through active mode-locking. This achievement was made possible by using a specific laser active medium, where the non-radiative upper state lifetime (and thus the gain recovery time) was artificially increased to several tens of picoseconds [278]. More recently, THz QCLs have been actively mode-locked by modulating their driving current with an external RF synthesizer [279], taking advantage of the longer non-radiative relaxation time of the upper laser state (~ 5-10 ps). The resulting 10 ps short pulses, limited in width by the sinusoidal microwave modulation, were detected using an indirect asynchronous sampling technique, exploiting an external mode-locked near-infrared laser.

Beside the inherent ultrafast gain recovery time, the spontaneous activation of mode-locking via saturable absorption in THz QCLs has traditionally faced a second issue: the total lack of consolidated saturable absorber technologies in this frequency range. However, graphene has allowed a major breakthrough in this field. Specifically, passive mode-locking in a THz QCL has been successfully demonstrated by integrating a distributed graphene saturable absorber (DGSA) on the top-surface of the double-metal QCL cavity (see Fig. 8b-c). This strategy enables the achievement of self-starting pulsed emission with 4.0-ps-long pulses (see Fig. 8d-g) in a compact, all-electronic, all-passive and inexpensive configuration [33]. The core idea takes advantage of the high transparency modulation (~80%) [67] and fast recovery time (2–3 ps) [69, 280] of graphene saturable absorption, which is faster than the gain recovery time in THz QCLs. In this scheme, the intracavity THz radiation experiences saturable losses, thereby promoting pulsed emission over the naturally occurring CW emission.

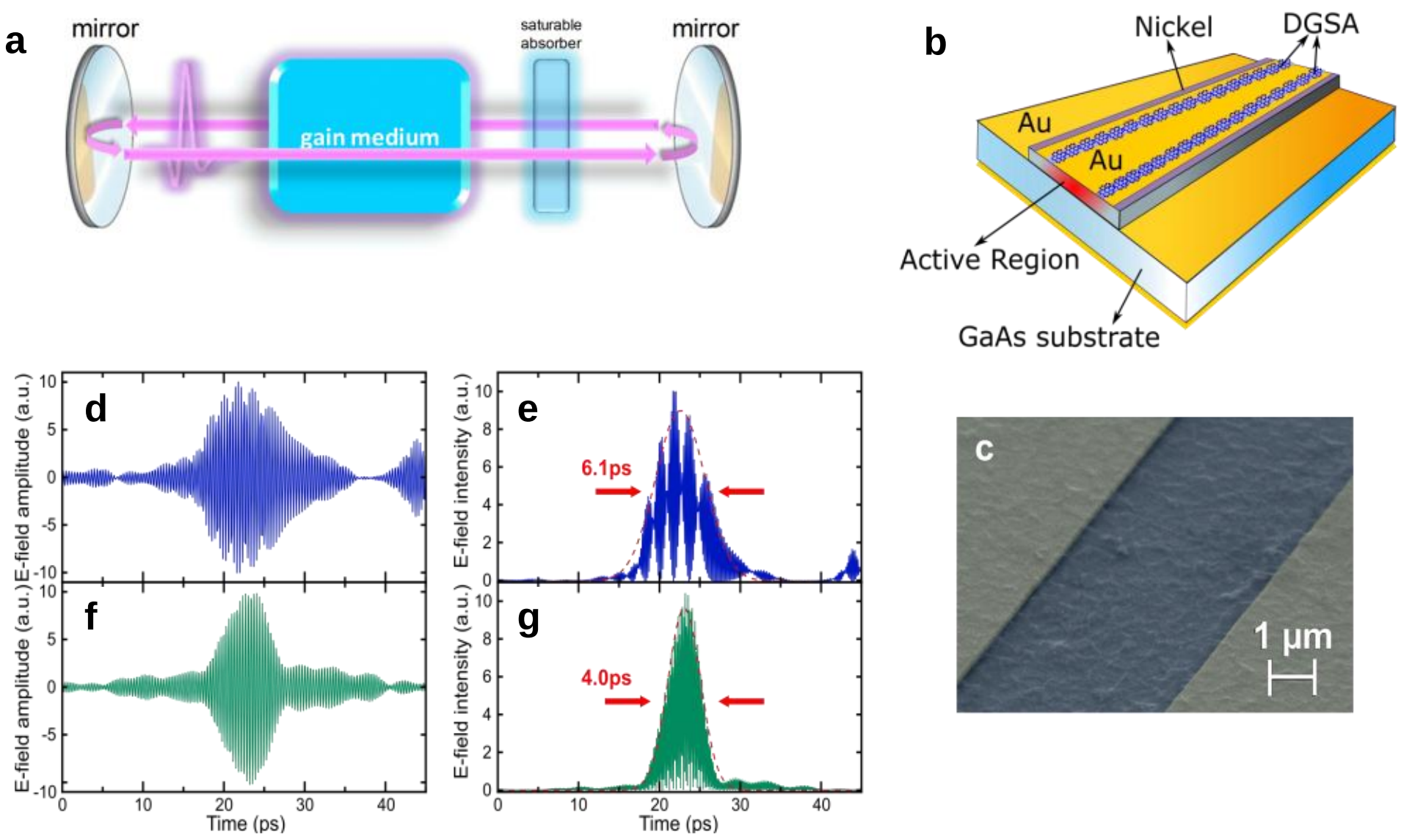

**Figure 8. Passive mode locking. (a)** Schematic diagram of a laser cavity with an intracavity saturable absorber (SA). The presence of the SA can favour the evolution and stabilization of short pulses travelling in the laser cavity. **(b)** Schematics of the device reported in ref. [33], featuring a distributed graphene saturable absorber (DGSA) integrated onto the metallic top contact of the double metal waveguide of a THz QCL. **(c)** False color scanning electron micrograph (SEM) image showing a portion of the QCL top surface, with the gold contact and the etched stripes covered by the multilayer graphene. (**d,e**) Electric field amplitude profile of a pulse emitted by the DGSA-QCL biased at 327 mA (d), along with its corresponding intensity profile (e). The dashed red line represents a Gaussian fit of the pulse with full-width at half-maximum (FWHM) of 6.1 ps. (**f,g**), Electric field amplitude profile of a pulse emitted by the DGSA-QCL at 570 mA (f), and corresponding intensity profile (g**)**. The dashed red line is a Gaussian fit of the pulse with FWHM of 4.0 ps. From Ref. [33]. Reprinted with permission from Nature Photonics 17, 607–614. Copyright 2023 Springer Nature Limited.

In ref. [33], multilayer graphene is patterned and shaped as a pair of stripes (Fig. 8c) integrated on the QCL top-contact. This introduces an additional loss channel to the QCL waveguide. Importantly, this loss is non-linear, and decreases for larger intracavity electric field strengths. Consequently, loss saturation creates a net gain window for the waves propagating along the laser cavity, facilitating pulse formation. Since graphene is distributed continuously along the entire cavity, this effect remains unaffected by the 2-3 ps fast gain recovery time of the investigated THz QCL active region, as it would happen if graphene was placed at a single point along the QCL cavity. The position and width of the DGSA stripes are carefully chosen as a trade-off between ensuring sufficient overlap with the electric field and avoiding excessive losses with respect to the reference QCL. The graphene stripes enable fast saturable absorption along the laser cavity. Additionally, the gain has a recovery time that is much faster (a few ps) than the roundtrip time [281]. As a result, simultaneous saturation of the gain and bleaching of the absorption occur as a consequence of the propagating pulse. A short net gain window is created, stabilizing the pulse and introducing, at the same time, a net loss for counterpropagating fields, ultimately suppressing the buildup of a CW background.

### 5.2 Frequency Combs

Frequency comb (FC) synthesizers are coherent optical sources that emit a spectrum consisting of a set of discrete, equally spaced optical modes with a well-defined phase relationship.

At THz frequencies, supercontinuum generation of FC has traditionally been achieved by optically rectifying mode-locked pulsed lasers, resulting in a power output of a few microwatts. THz QCLs can spontaneously operate as frequency combs [273, 282, 283], opening up new possibilities in previously unexplored domains, such as metrology, spectroscopy, and frequency synthesis. In a THz QCL, FC operation results from four-wave-mixing, which is a consequence of the third-order nonlinearity of the active medium. Indeed, the GaAs/AlGaAs heterostructures composing the active material have a large third-order $\chi^{(3)}$ susceptibility ($7 \times 10^{-16}$ $(m/V)^2$) [284], more that 6 orders of magnitude smaller than that of graphene. Specifically, the resonant Kerr nonlinearity can induce self-phase locking through an intracavity four-wave-mixing process (FWM) [273]. FWM tends to homogenize the mode spacing and, as a result, acts as the primary mechanism for mode proliferation and comb generation in a free-running QCL.

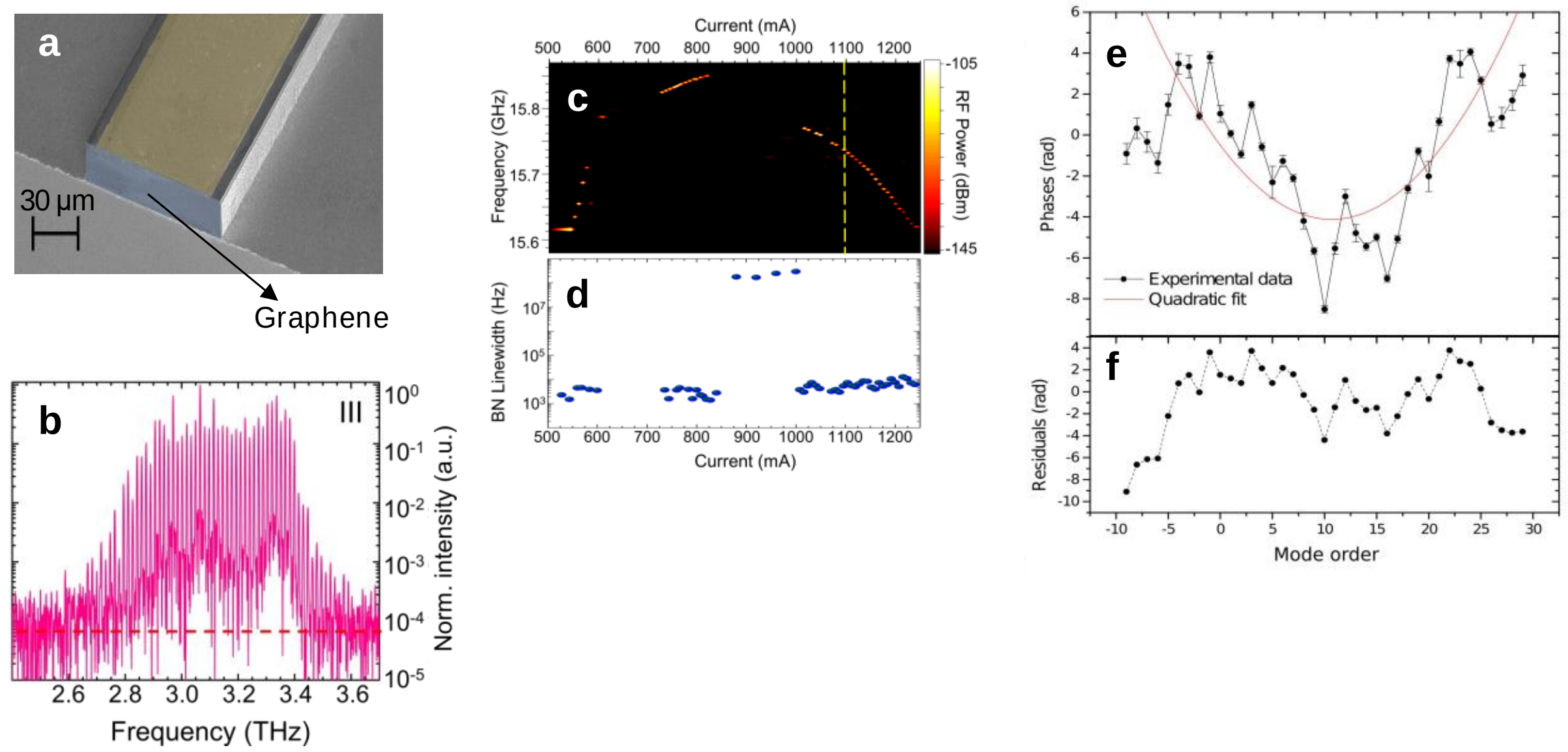

**Figure 9. Frequency combs. (a)** False color SEM image of the back facet of a QCL double metal waveguide coated with multilayer graphene. **(b)** Fourier transform infrared (FTIR) spectrum collected while driving the QCL in continuous wave at 1.1 A, corresponding to the driving current that maximizes the laser optical power. The dashed line indicates the noise level. **(c)** Intermode beatnote map and **(d)** beatnote linewidths, plotted as a function of the QCL current. The dashed yellow line indicates the point at which the metrological characterization has been performed. **(e,f)** Analysis of the Fourier phases of each mode of the QCL comb. Parabolic fit of the data and corresponding residuals are shown, resulting in a measured group delay dispersion GDD of (3.25 ± 0.5) $ps^2\ rad^{-1}$. From Ref. [285]. All panels (a-f) are from E. Riccardi et al., Laser Photonics Rev. 17, 2200412 (2023), licensed under a Creative Commons Attribution (CC BY) license.

However, in a free-running THz QCL, the FC regime is self-starting only if the group velocity dispersion [286-288] in the active medium is low enough to allow the cavity modes to be locked by the modes generated by FWM without the need of any additional optical elements. Therefore, it is essential to adapt the bias-dependent contribution to the dispersion while also maintaining and enhancing the optical bandwidth to achieve reliable control of the comb states across the entire operational range. To this end, gain medium engineering has proven to be a valuable solution for bandwidth optimization, but it significantly influences the dispersion dynamics at different bias points. As a result, other strategies have traditionally been adopted to compensate the group delay dispersion (GDD), including intracavity-integrated dispersion compensators [282, 286] or Gires–Tournois etalons. Recent works have demonstrated the effectiveness of these methods using various device configurations. For example, a planar mirror [289, 290], a graphene modulator [29], or a saturable absorber graphene reflector [291], individually placed behind the laser facet have been used to induce wavelength-dependent group delay, providing stable FC operation over 29%-55% of the laser operational range [29, 290-292]. Alternatively, *dc*-biased external cavity sections can compensate the dispersion, enabling FC operation over the entire QCL bias range, although over a relatively limited optical bandwidth (0.4 THz) [293]. Facet-integration of a non linear element, as graphene, in a Fabry-Perot QCL cavity can lead to a significant increase of the mirror losses with intensity discontinuity near the facet. As a consequence, this enhances the *cross-steepening* [294] terms between counter-propagating waves, which helps comb formation by increasing the intensity slope with respect to the distance as they propagate in the cavity.

A successful implementation of the above concept has recently been achieved by patterning multilayer graphene on the facet of a THz QCL (Fig. 9a). After light reflection from the graphene-covered facet, the THz field experiences positive feedback of its phase as it sees its phase at an earlier time in the counter-propagating wave. This influences the FC operation, resulting in a proliferation of emitted modes (Fig. 9b) over the entire gain bandwidth and across more than 60% of its operational range (Fig. 9c-d), with optical power ∼0.18 mW/mode [285], as predicted theoretically [295, 296], and a large phase coherence (Fig. 9e-f).

It is worth mentioning that the facet-integration of graphene can result in intensity-dependent losses in the external laser cavity [67, 249]. The reflection on the graphene surface contributes to the stabilization of the frequency comb through the same mechanism as the fast saturable gain in the active region of the QCL, thereby forcing it to frequency-modulated operation. Additionally, the saturable absorption effect observed in the multilayer graphene film also helps to regulate the remaining amplitude modulation, which can combine with the dominant frequency modulation effect [291].

### 5.3 Random Lasers

Random lasers (RLs) [297,298] are characterized by a random distribution of light scattering elements embedded in an active medium. These scatterers provide the required refractive index change and create a feedback mechanism that leads to light amplification by stimulated emission. Photons propagating in this *effective* cavity can undergo amplification and interact coherently multiple times in the gain medium. As a result, rich interference patterns are formed, characterized by an inherently high temporal coherence due to the long interaction length associated with the photon random walk in the disordered medium, and by a low spatial coherence [299], due to the coexistence of randomly distributed modes with distinct and spatially separated wave fronts.

The complexity of mode interaction in RLs, combined with the inherently large parameter space that governs their design [300], make RLs an ideal device platform for investigating exotic phenomena in photonics, including chaos [301], non-Gaussian statistics [302], non-Gaussian complexity [303], Anderson localization [304], and event the contro-intuitive mode-locking [305]. Understanding mode-locking in a RL requires detailed knowledge of the strength of the correlation mechanism and the number of cross-correlated interacting optical modes involved. A preliminary intuitive argument is that a spatial overlap between the modes is necessary [306]. While this provides a rough guideline for defining the main design parameters, it does not represent a sufficient condition to achieve mode-locking. Indeed, achieving mode correlation and locking also requires control over the interaction coupling strength between the different modes in a RL.

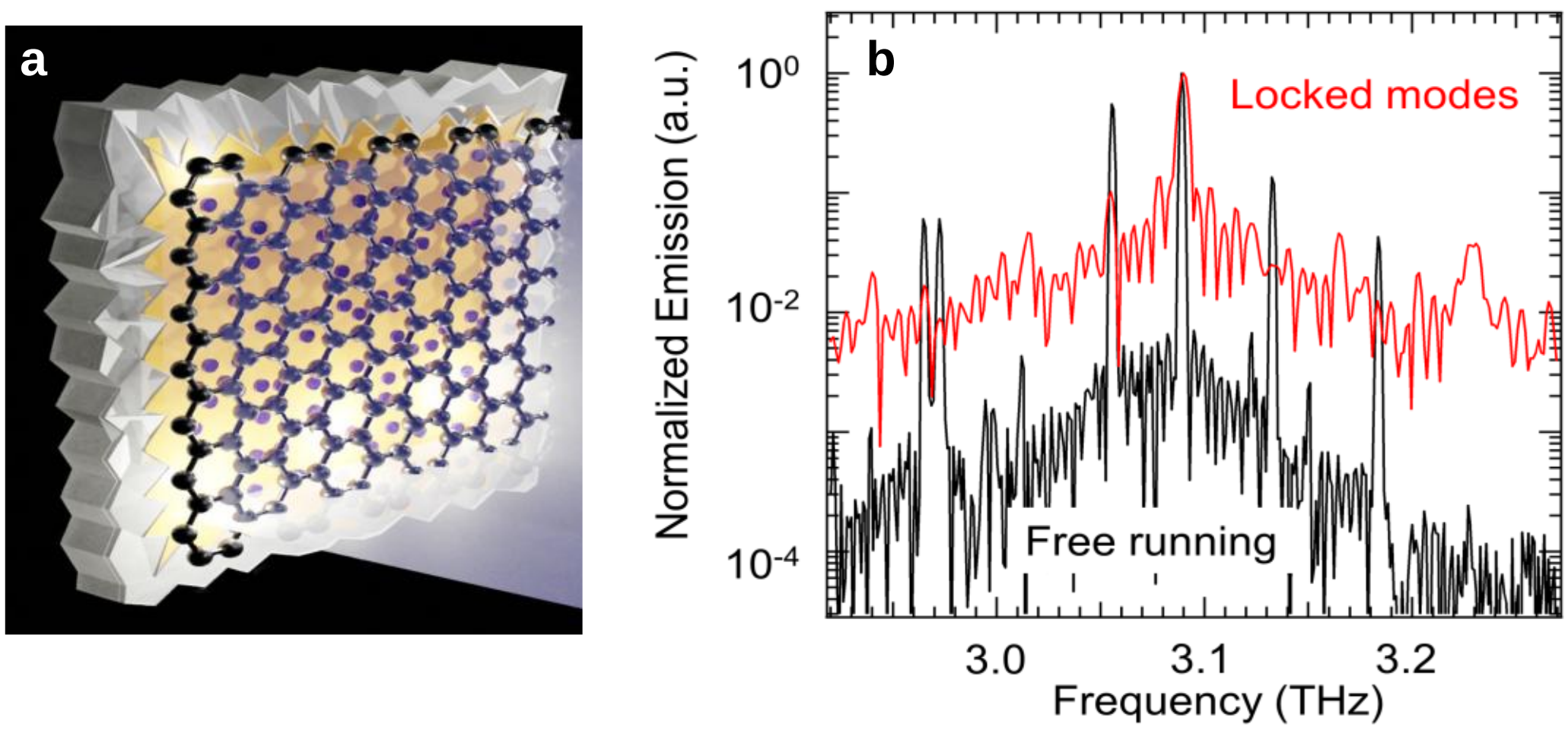

**Figure 10. Random lasers**. **(a)** Schematic representation of the graphene-integrated random QCL. The surface emitting laser, with a total area ~0.076 $mm^2$ (filling fraction 10%), comprises a bi-dimensional arrangement of $N = 99$ hole-scatterers, randomly distributed on a square area with size $L = 275$ µm, and irregular shaped borders coated with a 10-nm-thick layer of lossy Cr to suppress geometric cavity resonances. The photon mean free path for the random modes is 10.4 µm. **(b)** Comparison of normalized emission spectra, acquired via FTIR (black line) and by self-mixing intermode betanote (SMIB) spectroscopy (red line). From Ref. [307]: A. Di Gaspare et al., Adv. Sci. 10, 2206824 (2023), licensed under a Creative Commons Attribution (CC BY) license.

Although spontaneous mode-locking in RLs was theoretically predicted over a decade ago [305, 308], it has only recently been demonstrated at THz frequencies using graphene in surface emitting random THz QCLs [307]. By modifying the reflectivity of the RL top surface it is possible to alter the intracavity field of the THz QCL [309]. For this purpose, graphene has been used in two different configurations. In the first case, an ultrafast (~ps) graphene-based saturable absorber mirror [67] is coupled on-chip to the top surface of a random resonator. In the second case [307], multilayer (7 layers) graphene is lithographically inserted into the holes of a set of randomly arranged scatterers acting as light outcouplers (Fig. 10a). The analysis of comb operation was performed using self-mixing intermodal beatnote (SMIB) spectroscopy [307] (Fig. 10b), which unveiled, for the first time, the occurrence of mode-locking. The achievement of locking between spatially incoherent modes and their down-conversion to the radiofrequency (RF) domain, open important perspectives of use of random lasers for arbitrary frequency generation and reservoir computing in the far-infrared, marking a fascinating breakthrough in the physics of complex systems.

### 5.4 THz Third harmonic generation in graphene and graphene resonators

As previously discussed, graphene exhibits a remarkably high third-order THz susceptibility $\chi^{(3)}$~$10^{-9}$ $m^2/V^2$[73, 77, 310]. High harmonic generation (HHG), up to the 7th harmonic, even with moderate fields (~10 kV/cm) at frequencies ≤ 2.2 THz [78, 79, 84] have been reported in graphene, with field conversion efficiencies of $10^{-3}$, $10^{-4}$, and $10^{-5}$ for the third, fifth and seventh harmonics, respectively. This property greatly enhances non-linear interactions by orders of magnitude [73, 77, 310], enabling frequency 'up-conversion', provided that an incident beam of suitable power is used [78, 311], and/or through the use of metallic grating-based metamaterials [311].

The primary reason for the exceptionally efficient generation of THz high harmonics in graphene lies in the unique [312, 313] combination of quasi-instantaneous (~100 fs) carrier-carrier thermalization and THz-rate (~ps) cooling timescales to the driving electric fields.

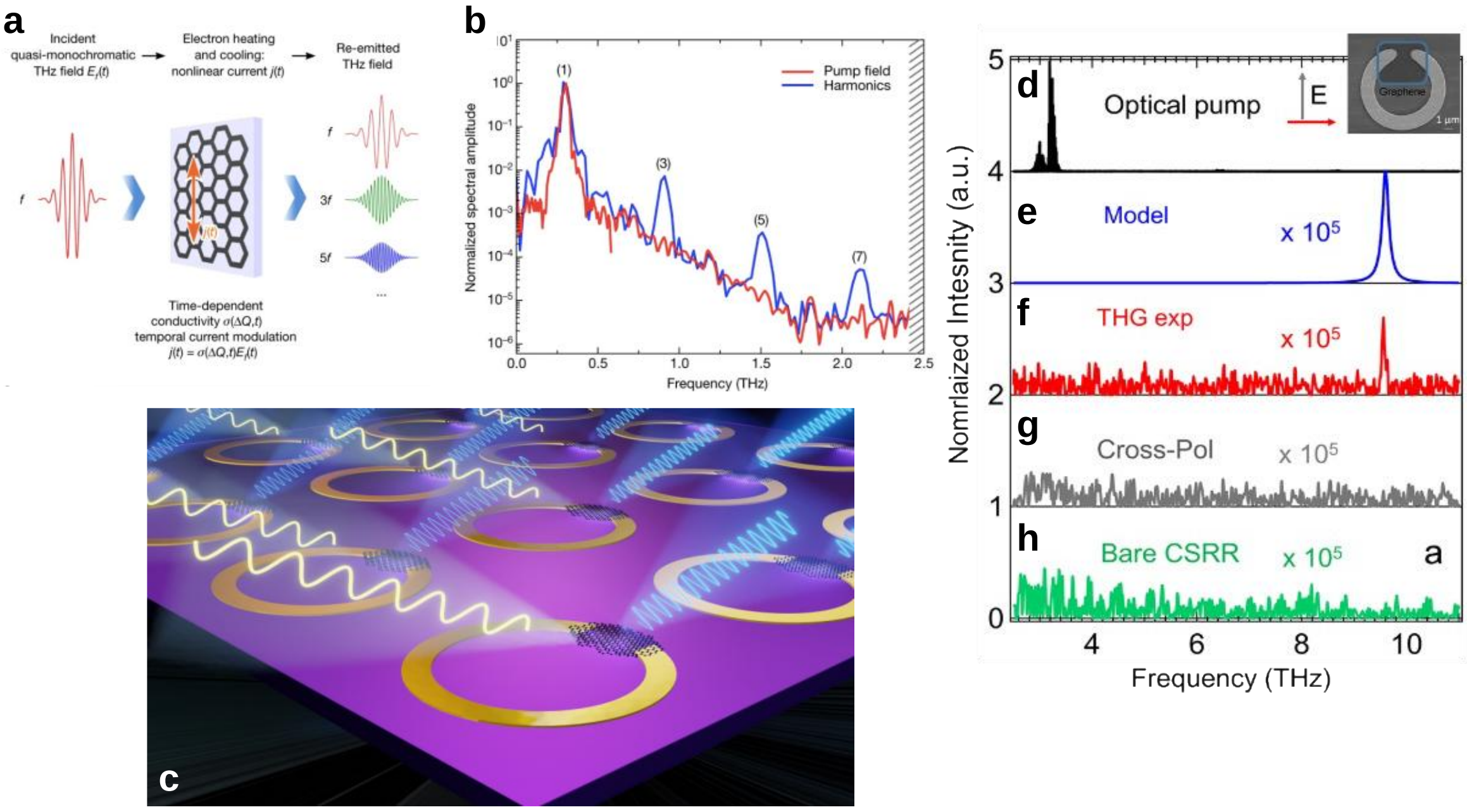

**Figure 11. High harmonic generation (a)** Schematic representation of high-harmonic generation in SLG. A quasi-monochromatic, linearly polarized THz wave (pump) is incident on a graphene sample, driving the nonlinear current in the graphene, and leading to re-emission at higher odd-order harmonics. From Ref. [78]: reprinted with permission from Nature 561, 507–511 (2018). Copyright 2018 Springer Nature Limited. **(b)** Red line: normalized amplitude spectrum of the incident pump THz wave at the fundamental frequency $f$ = 0.3 THz with peak field strength E = 85 kV $cm^{-1}$. Blue line: normalized spectrum of the same THz wave transmitted through graphene. From Ref. [78]: reprinted with permission from Nature 561, 507–511 (2018). Copyright 2018 Springer Nature Limited. **(c)** Third-harmonic generation in an array of split-ring-coupled single layer graphene. Schematics of the experiment. An array of CSRRs with an integrated monolayer graphene flake is optically pumped by a THz QCL, generating an optical signal at the third harmonic of the pumping beam. **(d-h)** Results of the experiment shown in (c). From top to bottom. (d) Normalized emission spectra of the THz QCL acquired at 15 K (e) Calculated THG efficiency for an input power $P_{exp}$=1.8 W. (f,g) Emission spectrum measured on the optically pumped SLG-CSRR, after filtering the QCL pump with a Ta-filter, with the SRR array oriented parallel (f)and perpendicular ( g) to the polarization axis of the QCL. (h) Emission from the an optically pumped metallic

CSRR array sample, without graphene, with the SRR array oriented parallel to the polarization axis of the QCL. Inset: scanning electron microscope image of the fabricated device. The inset of panel (d) shows the linear polarization directions employed to acquire the THG emissions: red curve (red arrow, active direction) and the gray curve (gray arrow, cross polarization) of the pump beam with respect to the CSRR array orientation. From ref. [318]: A. Di Gaspare et al., Nat. Commun. 15, 2312 (2024), licensed under a Creative Commons Attribution (CC BY) license.

.

THz third harmonic generation has also been demonstrated in topological insulators (TIs) [312, 313] and Dirac semi-metals [314, 315]. Importantly, it is possible to electrically tune harmonic generation [79], and to utilize ultrathin metamaterials that specifically increase harmonic power through field enhancement in a metallic grating located near the nonlinear Dirac-fermion system [311313]. Recently, this approach has achieved a record-high third-order susceptibility above $10^{-8}\,m^2\,V^{-2}$ in the THz regime for a grating-graphene metamaterial [311], exceeding the highest value obtained to date in a semiconductor quantum well [316, 317] by more than three orders of magnitude. To comprehend graphene impressive nonlinear performance, it is worth comparing its effective nonlinear optical coefficients for the third, fifth and seventh harmonics, which surpass those of typical solids by 7-18 orders of magnitude.

The demonstration of high harmonic generation in Dirac like materials at THz frequencies [78] often require complex facilities, such as free electron lasers, or large table-top laser systems as pump sources, thus limiting their applicability. Furthermore, the low conversion efficiency at frequencies above 2 THz hinders their use in applications within this frequency range.

Recently, emission at 9.63 THz in single layer graphene has been achieved using a compact geometry, by optically pumping an array of SLG coupled to circular split ring resonators with a high-power QCL with a central frequency of 3.21 THz (see Figure 11c) [318]. In this configuration, the mode confinement provided by the CSRR is essential to enhance the pump power density. This enables third harmonic generation (THG) in SLG in a frequency range (>3 THz, Figure 11d-h) so far unexplored.

## 6. Conclusions and Perspectives

In this review, we have gathered the main aspects of optoelectronic and photonic devices based on graphene and operating at THz frequencies. We have highlighted four main material properties that have guided the impressive developments of graphene-based THz devices: (i) strong infrared absorption due to a large intraband conductivity, (ii) ultra-fast carrier dynamics that can lead to intrinsically fast optoelectronic components, (iii) non-linear optical response to intense fields, enabling the realization of tailored emitters in the far-infrared, and (iv) electrical tunability of the aforementioned properties. In addition to these unique features, graphene represents a potentially disruptive material platform for its scalability and its potential for integration with pre-existing metallic- or semiconductor- based architectures. It is probably the only material that can be seamlessly integrated onto semiconductor THz lasers, large area metasurfaces, and photodetector focal plane arrays.

By examining at the four technological domains covered in this review, i.e. photodetectors, modulators, sources and non-linear devices, it is evident that graphene will impact differently on these four fields. In the context of photodetectors, graphene faces stiff competition from more established and mature technological platforms, such as superconducting bolometers, CMOS/microbolometer THz cameras and ultrafast Schottky diodes, making it challenging to develop receivers with unique features. Nonetheless, some noteworthy advancements have been obtained by using high quality LMHs to achieve extraordinary physical properties, underpinning the development of ultrabroadband [110] or polarization sensitive [152] devices operating at low temperatures. However, the widespread adoption of these solutions is hindered by the difficulty of scaling up high quality LMHs through technologically relevant fabrication workflows. A step change in this direction has been recently demonstrated with the creation of scalable large area heterostructures [116]. Despite this progress, the quality of graphene grown by chemical vapour deposition processes, is still not matching the superior performances of mechanically exfoliated materials. Therefore, the technological maturity of graphene-based photodetectors hinges on successfully navigating this complex, yet necessary, transition in the production of the active material.

In the context of THz modulators, large-area graphene has the potential to play a significant role in the long run compared to other materials like liquid crystals, vanadium dioxide, or silicon. While material quality is less critical for device performance in THz modulators than in photodetectors, many design concepts have been demonstrated but have yet to be implemented experimentally. For instance, the generation of THz OAM beams using graphene-based metasurfaces has been extensively investigated numerically, but there is still a lack of compelling device evidence [235].

As an innovative perspective, we feel that the development of integrated devices, such as hybrid QCLs-graphene emitters can be instrumental to enable novel and unforeseen functionalities in QCLs, as frequency tuning [209], self-starting mode-locking [285] and short pulse generation [33], potentially representing an enabling technology for the realization of future quantum emitters in the far-infrared. Additionally, graphene exhibits the strongest non-linear response in the THz frequency range, enabling the generation of radiation in challenging frequency ranges, such as ~7 - 10 THz, where the Reststrahlen band of GaAs-based heterostructures, constituting the active media of QCLs, hampers the realization of powerful and compact emitters. Table II summarize the state of the art of the technologies presented in the present review article.

| THz Technology | Material platform | Figure of merit | Relevance | Ref. |
|---|---|---|---|---|
| Room temperature photodetectors | hBN-encapsulated SLG | NEP < 100 pWHz$^{-1/2}$<br>Response time ~100 ps<br>Broadband operation (0.1 – 5 THz) | State-of-the-art combination in single-pixel | [112, 129] |
| | Large-area CVD SLG | NEP ~1 nWHz$^{-1/2}$<br>Response time ~1 ns<br>Broadband operation (0.1 – 5 THz) | Possibility of integration in large arrays | [27] |
| Low Temperature Photodetectors | SGS junction bolometer | NEP ~30 zWHz$^{-1/2}$ | Ultra-sensitive detector (towards single photon) | [118] |
| | Twisted double bilayer graphene | > 100 THz frequency coverage | Ultrabroadband photodetector from THz to mid-IR | [110] |
| Amplitude modulators | Graphene-based metamaterial | MD >80%<br>Speed >3 GHz | THz free-space communication | [202] |
| Reflectarrays | Antenna-integrated CVL SLG | Deflection angle ~25° | Beam steering and beam forming | [233] |
| Sources | QCL integrated DGSA | 4 ps pulse duration | First demonstration of passive mode-locking in a THz QCL | [33] |
| | QCL-facet integrated SLG | FC operation over >60% laser operational range | Dispersion compensation with direct integration of graphene | [285] |
| High-Harmonic Generation | CSRR-coupled graphene | 9.63 THz emission from a QCL | First demonstration of frequency up-conversion from a THz QCL | [318] |

**Table II.** Most recent device advancements described in the manuscript.

**Acknowledgements.** We acknowledged funding from the European Research Council (SPRINT) (681379) and Terascan (101157731), from the European Union through the FET Open project EXTREME IR (944735), the Graphene Flagship (core 3), the HORIZON-CL4-2021-DIGITAL-EMERGING-01 (101070546) Muquabis and the QuantERA 2021 project QATACOMB (Project No. 491 801 597) and from the European Union European Union under the Italian National Recovery and Resilience Plan (NRRP) of Next Generation EU, PNRR project PE0000023-NQSTI.

**Data availability:** The data presented in this study are available on request from the corresponding author.